\documentclass{article}%
\usepackage{amssymb}
\usepackage{amsmath}%
\usepackage{amsfonts}%
\usepackage{graphicx}
\usepackage{epstopdf}

\begin{document}

\title{A New Continuum Formulation for Materials--Part II. Some Applications in Fluid Mechanics}
\author{Melissa Morris\\3420 Campus Blvd. NE\\Albuquerque, NM 87106\\mmorris400@comcast.net}
\date{January 18, 2017}
\maketitle

\begin{abstract}
In \textsc{part I} of this paper, I proposed a new set of equations, which I
refer to as the $\overline{M}\left(  D,\eta\right)  $-formulation and which
differs from the Navier-Stokes-Fourier description of fluid motion. \ Here, I
use these equations to model several classic examples in fluid mechanics, with
the intention of providing a general sense of comparison between the two
approaches. \ A few broad facts emerge: (1) it is as simple--or in most cases,
much simpler--to find solutions with the $\overline{M}\left(  D,\eta\right)
$-formulation, (2) for some examples, there is not much of a difference in
predictions--in fact, for sound propagation and for examples in which there is
only a rotational part of the velocity, my transport coefficients $D$ and
$\eta$ are chosen to match Navier-Stokes-Fourier solutions in the appropriate
regimes, (3) there are, however, examples in which pronounced differences in
predictions appear, such as light scattering, and (4) there arise, moreover,
important conceptual differences, as seen in examples like sound at a
non-infinite impedance boundary, thermophoresis, and gravity's effect on the atmosphere.

\end{abstract}
\tableofcontents
\listoffigures
\listoftables

\section{\label{secprobs}Introduction}

In choosing the examples that appear in this paper, I desired the simplest
mathematics and closed-form solutions wherever possible. \ To this end, (1)
variation is limited to one spatial dimension in all sections but \S \ref{flh}
on hydrodynamical fluctuations and \S \ref{poise} on Poiseuille flow, (2)
equations are linearized about a constant state in all sections but
\S \ref{shock} on shock waves, and (3) parameters are chosen to yield
phenomena strictly in the hydrodynamic (small $K\!n$) regime in all sections
but \S \ref{shock}. \ Furthermore, so as not to distract from the main
results, for examples that require any mathematical complexity, such as the
stochastical subjects of \S \ref{flh} and \S \ref{ls}, I provide guiding
references and outline steps, but for the most part I merely present and
discuss the solutions. \ I will treat these problems--and others that are
examined only superficially here, like stability and sound propagation--with
rigor in future papers.

In addition, these particular examples were chosen to exhibit a wide variety
of phenomena--including ones that connect the two theories and so provide
anchoring points to tie $\overline{M}\left(  D,\eta\right)  $-formulation
parameters to those of Navier-Stokes-Fourier (NSF), and ones that reveal major
differences both measurably in experiment and conceptually.

The following is a brief description of the examples this paper contains.

\begin{itemize}
\item A one-dimensional linear stability analysis is carried out in
\S \ref{stability} to show that my formulation is unconditionally stable
provided that the diffusion parameter, $D$, is positive.

\item In \S \ref{diffusion}, a mass equilibration problem with no mechanical
forces is studied in order to show that my formulation reduces to one of pure
diffusion governed by Fick's law.

\item Low amplitude sound propagation, which is the subject of \S \ref{sound},
is demonstrated to be a way of relating my longitudinal diffusion parameter,
$D$,\ to the transport parameters of the NSF formulation by matching
attenuations in the hydrodynamic regime. \ I employ this relationship to
estimate values of $D$\ for several different gases and liquids, and these are
presented in \textsc{appendix \ref{difp}}.

\item In \S \ref{soundbc}, acoustic impedance at a sound barrier is discussed.
\ In particular, I point out that the $\overline{M}\left(  D,\eta\right)
$-formulation allows an impermeable boundary with a non-zero normal velocity
at that boundary, as that which occurs in the case of non-infinite impedance.

\item As a step towards bridging the gap between microscopic and macroscopic
scales in my theory, the subject of hydrodynamical fluctuations is addressed
in \S \ref{flh}, where it is shown that the thermodynamic interpretation of a
continuum mechanical point (i.e. a point occupied by matter and moving with
velocity $\underline{v}$) stemming from the $\overline{M}\left(
D,\eta\right)  $-formulation is fundamentally different from that of
Navier-Stokes-Fourier. \ That is, in the NSF theory, a continuum mechanical
point is viewed as a thermodynamic subsystem in contact with its surrounding
material which acts as a heat reservoir; whereas in my theory, the surrounding
material acts not only as a heat--but also a particle--reservoir.

\item The light scattering spectra from the $\overline{M}\left(
D,\eta\right)  $-formulation and the NSF equations are provided in
\S \ref{ls}. \ Comparing these, one finds the shifted Brillouin peaks, which
correspond to the sound (phonon) part of the spectrum, are virtually
indistinguishable between the two theories, as one would expect, but the
central Rayleigh peak predictions are significantly different. \ For example,
in a classical monatomic ideal gas in the hydrodynamic regime, the
$\overline{M}\left(  D,\eta\right)  $-formulation predicts a Rayleigh peak
that is about $29\%$ taller and narrower than the NSF equations (but with the
same area so that the well-verified Landau-Placzek ratio remains intact).
\ Previous experimenters, e.g. Clark \cite{clark} and Fookson et al.
\cite{fookson}, did not study gases fully in the hydrodynamic regime,
$K\!n\lesssim O\left(  10^{-2}\right)  $, choosing instead to focus on more
rarefied gases with Knudsen numbers in the range, $O\left(  10^{-1}\right)
\lesssim K\!n\lesssim O\left(  1\right)  $,\ which extends into the slip flow
and transition regimes. \ Therefore, in order to verify the $\overline
{M}\left(  D,\eta\right)  $-formulation, it is important to conduct a
high-resolution light scattering experiment for a gas in the hydrodynamic regime.

\item In \S \ref{grav}, I study the effect of gravity on the Earth's lower
atmosphere, and show that enforcing no mass flux and no total energy flux
conditions in the $\overline{M}\left(  D,\eta\right)  $-formulation, leads to
the isentropic condition that is typically\ assumed a priori. \ In contrast,
enforcing the same conditions in the NSF formulation, leads to an isothermal
condition, which is obviously not physical.

\item Poiseuille flow is the subject of \S \ref{poise}. \ There, I show that,
when compared to the NSF mass flow rate, which is due solely to convection,
the $\overline{M}\left(  D,\eta\right)  $-formulation predicts an additional
contribution due to diffusion. \ In the hydrodynamic regime, however, this
contribution is very small,\ and so these predictions do not differ by much.
\ Since viscometers may be based on Poiseuille flow, this--and other examples
of this sort--provide justification that the shear viscosity appearing in the
$\overline{M}\left(  D,\eta\right)  $-formulation may be taken to be the same
as the Navier-Stokes shear viscosity.

\item Thermophoresis, or the down-temperature-gradient motion of a macroscopic
particle in a resting fluid, is discussed in \S \ref{therm}. \ For an
idealized problem, it is shown that, through the steady-state balancing of the
convective and diffusive terms of the mass flux, the $\overline{M}\left(
D,\eta\right)  $-formulation provides a mechanism for thermophoresis that is
not present in the NSF formulation. \ In the latter type, a thermal slip
boundary condition at the particle's surface is needed in order to model
thermophoretic motion, whereas the former type may be used to describe this
motion with a no-slip boundary condition, i.e. one in which the tangential
velocity of the particle at its surface equals that of the fluid. \ The
concept of thermal slip is based on kinetic gas theory arguments in the slip
flow regime, appropriate for $K\!n\gtrsim O\left(  10^{-2}\right)  $, yet
thermophoresis is observed in gases for much smaller Knudsen numbers in the
hydrodynamic regime and also in liquids. \ Therefore, there are obvious
advantages to being able to model this problem without the thermal slip condition.

\item A non-linear steady-state shock wave problem is considered in
\S \ref{shock}. \ It is well-known that for this problem, the NSF
formulation--and all other accepted theories, for that matter--produces
normalized density, velocity, and temperature shock wave profiles that are
appreciably displaced from one another with the temperature profile in the
leading edge of the shock, velocity behind it, and density trailing in the
back. \ The $\overline{M}\left(  D,\eta\right)  $-formulation, however,
predicts virtually no displacement between these three profiles in the leading
edge of the shock and much less pronounced displacements in the middle and
trailing end of the shock.
\end{itemize}

\section{Equations of Motion from Part I}

All formulas referenced from \textsc{part I}\ of this paper are labeled with
"I." preceding their number. \ As derived in \textsc{part I}, the
$\overline{M}\left(  D,\eta\right)  $-formulation consists of the balance laws
(I.43):%
\begin{align}
\frac{\partial\overline{m}}{\partial t}  & =-\nabla\cdot\left(  \overline
{m}\underline{v}\right) \nonumber\\
\frac{\partial m}{\partial t}  & =-\nabla\cdot\left(  \underline{q}%
_{M}+m\underline{v}\right) \label{s1}\\
\frac{\partial\left(  \overline{m}\underline{v}\right)  }{\partial t}  &
=-\nabla\cdot\left(  \underline{\underline{P}}+\overline{m}\underline
{v}\,\underline{v}\right)  +\overline{m}\underline{f}_{\overline{M}%
}\nonumber\\
\frac{\partial u}{\partial t}  & =-\nabla\cdot\left(  \underline{q}%
_{U}+u\underline{v}\right)  -\underline{\underline{P}}\colon\nabla
\underline{v}\nonumber
\end{align}
with constitutive equations (I.74):%
\begin{align}
\underline{q}_{M}  & =-D\nabla m\nonumber\\
\underline{\underline{P}}  & =P\underline{\underline{1}}+\underline
{\underline{P}}_{\text{visc}}\text{ with }\underline{\underline{P}%
}_{\text{visc}}=\left[
\begin{array}
[c]{c}%
-\left(  \overline{m}D-\frac{4\eta}{3}\right)  \left(  \nabla\cdot
\underline{v}\right)  \underline{\underline{1}}-\\
2\eta\left(  \nabla\underline{v}\right)  ^{sy,dev}%
\end{array}
\right] \label{s2}\\
\underline{q}_{U}  & =-D\nabla u.\nonumber
\end{align}

The Navier-Stokes-Fourier formulation is given by the balance laws (I.45):%
\begin{align}
\frac{\partial m}{\partial t}  & =-\nabla\cdot\left(  m\underline{v}\right)
\nonumber\\
\frac{\partial\left(  m\underline{v}\right)  }{\partial t}  & =-\nabla
\cdot\left(  \underline{\underline{P}}+m\underline{v}\,\underline{v}\right)
+m\underline{f}_{M}\label{s3}\\
\frac{\partial u}{\partial t}  & =-\nabla\cdot\left(  \underline{q}%
_{U}+u\underline{v}\right)  -\underline{\underline{P}}\colon\nabla
\underline{v}\nonumber
\end{align}
with constitutive equations (I.46):%
\begin{align}
\underline{\underline{P}}  & =P\underline{\underline{I}}+\underline
{\underline{P}}_{\text{visc}}\text{ with }\underline{\underline{P}%
}_{\text{visc}}=-\zeta_{NS}\left(  \nabla\cdot\underline{v}\right)
\underline{\underline{I}}-2\eta\left(  \nabla\underline{v}\right)
^{sy,dev}\label{s4}\\
\underline{q}_{U}  & =-k_{F}\nabla T.\nonumber
\end{align}

Other quantities employed in this paper that hold for both formulations when
the proper constitutive equations are used are\ the total energy balance law
(I.8)/(I.20),%
\begin{equation}
\frac{\partial e}{\partial t}=-\nabla\cdot\left(  \underline{q}_{U}%
+\underline{\underline{P}}\cdot\underline{v}+e\underline{v}\right)
,\label{s5}%
\end{equation}
the total energy flux,%
\begin{equation}
\underline{j}_{E}=\underline{q}_{U}+\underline{\underline{P}}\cdot
\underline{v}+e\underline{v},\label{s6}%
\end{equation}
and the total mass flux,%
\begin{equation}
\underline{j}_{M}=\underline{q}_{M}+m\underline{v}.\label{s7}%
\end{equation}

To solve the steady-state examples of \S \ref{grav}-\ref{therm}, one may
derive the following convenient forms for the two formulations considered
here. \ First, taking the balance laws (\ref{s1} a-c) and (\ref{s5}) with
constitutive equations (\ref{s2}), setting the time-derivatives equal to zero
for a steady state, using thermodynamic relationships (\ref{e4}) and
(\ref{e5}) to express the equations in terms of the variables $\overline{m}$,
$P$, $\underline{v}$, and $T$, and linearizing about a constant
state,\footnote{The subscript "$\ast$" is used to indicate that the quantity,
written as a function of $P$ and $T$, is evaluated at the thermodynamic state
$\left(  P_{\ast},T_{\ast}\right)  $.}%
\begin{equation}
\left(  \overline{m},P,\underline{v},T\right)  =\left(  m_{\ast},P_{\ast
},\underline{0},T_{\ast}\right)  ,\label{s8}%
\end{equation}
one obtains%
\begin{align}
0  & =-m_{\ast}\nabla\cdot\underline{v}\nonumber\\
0  & =D_{\ast}\left[  -\left(  m\alpha_{P}\right)  _{\ast}\nabla^{2}T+\left(
\frac{\gamma}{c^{2}}\right)  _{\ast}\nabla^{2}P\right]  -m_{\ast}\nabla
\cdot\underline{v}\label{s9}\\
0  & =-\nabla P+\left(  \overline{m}D-\frac{4\eta}{3}\right)  _{\ast}%
\nabla\left(  \nabla\cdot\underline{v}\right)  +2\eta_{\ast}\nabla\cdot\left[
\left(  \nabla\underline{v}\right)  ^{sy,dev}\right] \nonumber\\
0  & =D_{\ast}\left\{
\begin{array}
[c]{c}%
\left[  m\left(  c_{P}-h_{M}\alpha_{P}\right)  \right]  _{\ast}\nabla^{2}T+\\
\left(  \frac{h_{M}\gamma}{c^{2}}-T\alpha_{P}\right)  _{\ast}\nabla^{2}P
\end{array}
\right\}  -\left(  mh_{M}\right)  _{\ast}\nabla\cdot\underline{v}.\nonumber
\end{align}
If we then substitute (\ref{s9} a) into (\ref{s9} c-d) and employ the tensor
identity (\ref{to31r}), the above equations imply that for the $\overline
{M}\left(  D,\eta\right)  $-formulation,%
\begin{align}
0  & =\nabla\cdot\underline{v}\nonumber\\
0  & =\nabla^{2}P\label{s10}\\
0  & =-\nabla P+\eta_{\ast}\nabla^{2}\underline{v}+m_{\ast}f_{\overline{M}%
}\nonumber\\
0  & =\nabla^{2}T.\nonumber
\end{align}
In addition, by substituting constitutive relations (\ref{s2}) into equations
(\ref{s6}) and (\ref{s7}), employing thermodynamic relations (\ref{e4}),
(\ref{e5}), and (\ref{ee10}), and linearizing about constant state (\ref{s8}),
one finds the following total energy and mass fluxes for the $\overline
{M}\left(  D,\eta\right)  $-formulation:%
\begin{equation}
\underline{j}_{E}=-D_{\ast}\left\{
\begin{array}
[c]{c}%
\left[  m\left(  c_{P}-h_{M}\alpha_{P}\right)  \right]  _{\ast}\nabla T+\\
\left(  \frac{h_{M}\gamma}{c^{2}}-T\alpha_{P}\right)  _{\ast}\nabla P
\end{array}
\right\}  +\left(  mh_{M}\right)  _{\ast}\underline{v}\label{s11}%
\end{equation}
and%
\begin{equation}
\underline{j}_{M}=-D_{\ast}\left[  -\left(  m\alpha_{P}\right)  _{\ast}\nabla
T+\left(  \frac{\gamma}{c^{2}}\right)  _{\ast}\nabla P\right]  +m_{\ast
}\underline{v}\label{s12}%
\end{equation}
where, in addition to all of the quantities defined in \textsc{part I}, we
have introduced the isobaric specific heat per mass $c_{P}$, isobaric to
isochoric specific heat ratio $\gamma=c_{P}/c_{V}$, and adiabatic sound speed
$c$. \ Carrying out similar steps, yields for the NSF formulation, the
steady-state equations,%
\begin{align}
0  & =\nabla\cdot\underline{v}\nonumber\\
0  & =-\nabla P+\eta_{\ast}\nabla^{2}\underline{v}+m_{\ast}f_{M}\label{s13}\\
0  & =\nabla^{2}T,\nonumber
\end{align}
and the fluxes,%
\begin{equation}
\underline{j}_{E}=-\left(  k_{F}\right)  _{\ast}\nabla T+\left(
mh_{M}\right)  _{\ast}\underline{v}\label{s14}%
\end{equation}
and%
\begin{equation}
\underline{j}_{M}=m_{\ast}\underline{v}.\label{s15}%
\end{equation}
Notice that the set of equations (\ref{s13}) is identical to the set
(\ref{s10}), except for the absence of Laplace's equation for the pressure
(\ref{s10} b). \ Also, the fluxes (\ref{s14}) and (\ref{s15}) have the same
convective parts as the $\overline{M}\left(  D,\eta\right)  $-formulation
fluxes (\ref{s11}) and (\ref{s12}), but differing diffusive parts (with the
NSF mass flux having no diffusion at all).

\section{\label{stability}Stability Analysis}

Let us consider the Cartesian one-dimensional problem in which variation is
assumed to be in the $x_{1}\equiv x$ direction only with $v_{1}\equiv v$\ as
the only non-zero component of the velocity and there are assumed to be no
body forces. \ If we use the thermodynamic relationships (\ref{e.6}) and
(\ref{e.7}) to recast the $\overline{M}\left(  D,\eta\right)  $-formulation
(\ref{s1})/(\ref{s2}) in terms of the variables, $\overline{m}$, $m$, $v$, and
$T$, and then linearize about the constant equilibrium state,%
\begin{equation}
\left(  \overline{m},m,v,T\right)  =\left(  m_{\text{eq}},m_{\text{eq}%
},0,T_{\text{eq}}\right)  ,\label{c1}%
\end{equation}
via%
\begin{align}
\overline{m}  & =m_{\text{eq}}+\delta\!\overline{m}\label{c2}\\
m  & =m_{\text{eq}}+\delta\!m\label{c3}\\
v  & =\delta\!v\label{c4}\\
T  & =T_{\text{eq}}+\delta\!T,\label{c5}%
\end{align}
assuming%
\begin{align*}
\left\vert \delta\!\overline{m}\right\vert ,\left\vert \delta\!m\right\vert  &
\ll m_{\text{eq}}\\
\left\vert \delta\!T\right\vert  & \ll T_{\text{eq}}\\
\left\vert \delta\!v\right\vert  & \ll c_{\text{eq}},
\end{align*}
then we arrive at the following system of linear equations:%
\begin{align}
\frac{\partial\delta\!\overline{m}}{\partial t}  & =-m_{\text{eq}}%
\frac{\partial\delta\!v}{\partial x}\label{c6}\\
\frac{\partial\delta\!m}{\partial t}  & =D_{\text{eq}}\frac{\partial^{2}%
\delta\!m}{\partial x^{2}}-m_{\text{eq}}\frac{\partial\delta\!v}{\partial
x}\label{c7}\\
\frac{\partial\delta\!v}{\partial t}  & =-\frac{1}{\left(  m^{2}\kappa
_{T}\right)  _{\text{eq}}}\frac{\partial\delta\!m}{\partial x}-\left(
\frac{\alpha_{P}}{m\kappa_{T}}\right)  _{\text{eq}}\frac{\partial\delta
\!T}{\partial x}+D_{\text{eq}}\frac{\partial^{2}\delta\!v}{\partial x^{2}%
}\label{c8}\\
\frac{\partial\delta\!T}{\partial t}  & =D_{\text{eq}}\frac{\partial^{2}%
\delta\!T}{\partial x^{2}}-\left(  \frac{T\alpha_{P}}{m\kappa_{T}c_{V}%
}\right)  _{\text{eq}}\frac{\partial\delta\!v}{\partial x},\label{c9}%
\end{align}
where the subscript "eq" indicates that the parameter is evaluated at the
constant equilibrium state (\ref{c1}). \ Note that for this linearized
problem, the mechanical mass equation (\ref{c6}) may be uncoupled from the
rest of the system, (\ref{c7})-(\ref{c9}). \ Postulating a solution,%
\[
\left[
\begin{array}
[c]{c}%
\delta\!m\\
\delta\!v\\
\delta\!T
\end{array}
\right]  ,
\]
to (\ref{c7})-(\ref{c9}) that is proportional to%
\[
\exp\left(  i\kappa x+\omega t\right)  ,
\]
for $\kappa$ real and $\omega$ complex, one obtains the dispersion relation,%
\begin{equation}
\left[
\begin{array}
[c]{c}%
\omega^{3}+3D_{\text{eq}}\kappa^{2}\omega^{2}+\left(  c_{\text{eq}}%
^{2}+3D_{\text{eq}}^{2}\kappa^{2}\right)  \kappa^{2}\omega+\\
D_{\text{eq}}\left(  c_{\text{eq}}^{2}+D_{\text{eq}}^{2}\kappa^{2}\right)
\kappa^{2}%
\end{array}
\right]  =0.\label{c11}%
\end{equation}
In the above, I have employed the equilibrium thermodynamic relationships
(\ref{e11}) and (\ref{e12}). \ Equation (\ref{c11}) may be solved for $\omega$
to obtain the three exact roots,\footnote{Note that these roots being exact
enables us to construct exact Green's functions on the infinite domain.}%
\begin{align}
\omega_{1}\left(  \kappa\right)   & =-D_{\text{eq}}\kappa^{2}\label{c13}\\
\omega_{2}\left(  \kappa\right)   & =-D_{\text{eq}}\kappa^{2}+ic_{\text{eq}%
}\kappa\label{c14}\\
\omega_{3}\left(  \kappa\right)   & =-D_{\text{eq}}\kappa^{2}-ic_{\text{eq}%
}\kappa.\label{c15}%
\end{align}
Clearly, if%
\begin{equation}
D_{\text{eq}}>0\label{c16}%
\end{equation}
is satisfied, then the real parts of $\omega_{1}\left(  \kappa\right)  $,
$\omega_{2}\left(  \kappa\right)  $, and $\omega_{3}\left(  \kappa\right)  $,
are negative for all $\kappa$, resulting in the unconditional stability of
linearized system (\ref{c7})-(\ref{c9}).

For comparison, carrying out a similar procedure with the NSF formulation
(\ref{s3})/(\ref{s4}) yields the linearization,%
\begin{align}
\frac{\partial\delta\!m}{\partial t}  & =-m_{\text{eq}}\frac{\partial
\delta\!v}{\partial x}\label{c16.3}\\
\frac{\partial\delta\!v}{\partial t}  & =\left[
\begin{array}
[c]{c}%
-\frac{1}{\left(  m^{2}\kappa_{T}\right)  _{\text{eq}}}\frac{\partial
\delta\!m}{\partial x}-\left(  \frac{\alpha_{P}}{m\kappa_{T}}\right)
_{\text{eq}}\frac{\partial\delta\!T}{\partial x}+\\
\left(  \frac{\zeta_{NS}}{m}+\frac{4\eta}{3m}\right)  _{\text{eq}}%
\frac{\partial^{2}\delta\!v}{\partial x^{2}}%
\end{array}
\right] \label{c16.4}\\
\frac{\partial\delta\!T}{\partial t}  & =\left(  \frac{k_{F}}{mc_{V}}\right)
_{\text{eq}}\frac{\partial^{2}\delta\!T}{\partial x^{2}}-\left(  \frac
{T\alpha_{P}}{m\kappa_{T}c_{V}}\right)  _{\text{eq}}\frac{\partial
\delta\!v_{x,1}}{\partial x},\label{c16.5}%
\end{align}
with the dispersion relation,%
\begin{equation}
\left(
\begin{array}
[c]{c}%
\omega^{3}+\left[  \frac{\zeta_{NS}}{m}+\left(  \frac{4}{3}+E\!u\right)
\frac{\eta}{m}\right]  _{\text{eq}}\kappa^{2}\omega^{2}+\\
\left\{  c_{\text{eq}}^{2}+\left[  E\!u\left(  \frac{\zeta_{NS}}{m}%
+\frac{4\eta}{3m}\right)  \frac{\eta}{m}\right]  _{\text{eq}}\kappa
^{2}\right\}  \kappa^{2}\omega+\\
\left(  \frac{E\!u}{\gamma}\frac{\eta}{m}c^{2}\right)  _{\text{eq}}\kappa^{4}%
\end{array}
\right)  =0,\label{c17}%
\end{equation}
where $\gamma$ is the ratio of specific heats and $E\!u$\ is the Euken ratio
defined to be%
\begin{equation}
E\!u=\frac{k_{F}}{c_{V}\eta}.\label{c18}%
\end{equation}
Let us also define the quantities,%
\begin{equation}
\Sigma=\frac{E\!u}{\gamma}\frac{\eta}{m}\label{c18.1}%
\end{equation}
and%
\begin{equation}
\Gamma=\frac{1}{2}\left\{  \frac{\zeta_{NS}}{m}+\left[  \frac{4}{3}+\left(
1-\frac{1}{\gamma}\right)  E\!u\right]  \frac{\eta}{m}\right\}  .\label{c18.2}%
\end{equation}
Using these, equation (\ref{c17}) yields the following three approximate
roots,%
\begin{align}
\omega_{1}\left(  \kappa\right)   & \approx-\Sigma_{\text{eq}}\kappa
^{2}\label{c19}\\
\omega_{2}\left(  \kappa\right)   & \approx-\Gamma_{\text{eq}}\kappa
^{2}+ic_{\text{eq}}\kappa\label{c20}\\
\omega_{3}\left(  \kappa\right)   & \approx-\Gamma_{\text{eq}}\kappa
^{2}-ic_{\text{eq}}\kappa,\label{c21}%
\end{align}
for%
\begin{equation}
\left(  \frac{\eta}{mc}\right)  _{\text{eq}}\left\vert \kappa\right\vert
\ll1,\label{c22}%
\end{equation}
which corresponds to the low Knudsen number, or hydrodynamic, regime. \ For an
equilibrium thermodynamically stable fluid,%
\begin{equation}
\gamma\geq1\text{ and }c_{V}>0\label{c23}%
\end{equation}
are satisfied, and therefore, if the standard assumption that $\eta
_{\text{eq}},\left(  k_{F}\right)  _{\text{eq}}>0$ and $\left(  \zeta
_{NS}\right)  _{\text{eq}}\geq0$ is made, then the one-dimensional linearized
NSF formulation is stable in the hydrodynamic regime, as well.

In two future papers, \cite{M1d} and \cite{M3d}, I will examine linear
stability in more detail by constructing Green's functions on one and
three-dimensional infinite domains for the general $\overline{M}$-formulation,
which includes both the $\overline{M}\left(  D,\eta\right)  $ and NSF
formulations. \ In \cite{M3d},\ it will be shown that the three-dimensional
stability requirements on the transport parameters are%
\begin{equation}
D,\eta>0\label{cn1}%
\end{equation}
for the $\overline{M}\left(  D,\eta\right)  $-formulation and%
\begin{equation}
\eta,\left(  \frac{4}{3}\eta+\zeta_{NS}\right)  ,k_{F}>0\label{cn2}%
\end{equation}
for NSF. \ In another future paper \cite{stochIII}, I will demonstrate the
above criteria to be identical to those obtained by a particular version of
the second law of thermodynamics, which I argue should be used in place of the
version employed by de Groot and Mazur \cite[\textsc{ch. IV}]{degroot}.

\section{\label{diffusion}Pure Diffusion}

If we take the limit as the transport parameters go to zero ($D,\eta
\rightarrow0$ in the $\overline{M}\left(  D,\eta\right)  $-formulation or
$\eta,\zeta_{NS},k_{F}\rightarrow0$ in the NSF formulation), then we are left
with the Euler equations of pure wave motion for which, on an infinite domain,
perturbations travel at the adiabatic sound speed and never decay. \ However,
in the present section, we study a problem at the opposite extreme, one of
pure diffusion.

For this, it is instructive to bear the following example of thermodynamic
equilibration in mind. \ Let us consider a Cartesian one-dimensional problem
on the infinite domain in which the initial conditions are given by%
\begin{align}
m\left(  x,0\right)   & =\left\{
\begin{array}
[c]{l}%
m_{\text{eq}}+\Delta_{m}\text{ for }x<\left\vert \frac{L}{2}\right\vert \\
m_{\text{eq}}\text{ for }x>\left\vert \frac{L}{2}\right\vert
\end{array}
\right. \label{c23.01}\\
v\left(  x,0\right)   & =0\text{ for all }x\label{c23.02}\\
T\left(  x,0\right)   & =\left\{
\begin{array}
[c]{l}%
T_{\text{eq}}+\Delta_{T}\text{ for }x<\left\vert \frac{L}{2}\right\vert \\
T_{\text{eq}}\text{ for }x>\left\vert \frac{L}{2}\right\vert
\end{array}
\right.  .\label{c23.03}%
\end{align}
This describes a perturbed subsystem initially shaped like an infinite slab,
centered at $x=0$ with width $L$, in contact with two identical infinite
reservoirs on either side of it. \ As time approaches infinity, one expects
the subsystem to equilibrate with the reservoirs until the the whole system
has uniform mass density $m_{\text{eq}}$\ and temperature $T_{\text{eq}}$.
\ Let us further suppose (1) that the perturbations $\Delta_{m}$\ and
$\Delta_{T}$\ are small compared their respective equilibrium values so that
it is appropriate to use the linearized equations presented in
\S \ref{stability} and (2) that, in view of equilibrium thermodynamic relation
(\ref{e.7}), $\Delta_{m}$\ and $\Delta_{T}$\ are chosen to satisfy%
\begin{equation}
\Delta_{m}=-\left(  m\alpha_{P}\right)  _{\text{eq}}\Delta_{T}\label{c23.04}%
\end{equation}
so that the initial pressure is in equilibrium, i.e.%
\begin{equation}
P\left(  x,0\right)  =P_{\text{eq}}\text{ for all }x\text{.}\label{c23.05}%
\end{equation}
Note that conditions (\ref{c23.02}) and (\ref{c23.05}) mean that there are
initially no mechanical forces acting on the system, and so it is of interest
to see, in a problem such as this, the consequences of assuming a solution
having $v\left(  x,t\right)  =0$.

To this end, let us postulate the zero velocity solution,%
\begin{equation}
\delta\!v\left(  x,t\right)  =0,\label{c23.1}%
\end{equation}
to my linearized equations (\ref{c7})-(\ref{c9}). \ Substituting (\ref{c23.1})
into the equations yields%
\begin{equation}
\frac{\partial\delta\!m}{\partial t}=D_{\text{eq}}\frac{\partial^{2}\delta
\!m}{\partial x^{2}},\label{c23.2}%
\end{equation}%
\begin{equation}
\left(  \frac{1}{m\kappa_{T}}\right)  _{\text{eq}}\frac{\partial\delta
\!m}{\partial x}+\left(  \frac{\alpha_{P}}{\kappa_{T}}\right)  _{\text{eq}%
}\frac{\partial\delta\!T}{\partial x}=0,\label{c23.3}%
\end{equation}
and%
\begin{equation}
\frac{\partial\delta\!T}{\partial t}=D_{\text{eq}}\frac{\partial^{2}\delta
\!T}{\partial x^{2}}.\label{c23.4}%
\end{equation}
Equation (\ref{c23.3}) is a constraint that there are no pressure gradients
and it forces the mass density and temperature perturbations to be related by%
\begin{equation}
\delta\!m\left(  x,t\right)  =-\left(  m\alpha_{P}\right)  _{\text{eq}}%
\delta\!T\left(  x,t\right) \label{c23.5}%
\end{equation}
as long as there exists any point at which $m\ $and $T$\ attain their
equilibrium values, $m_{\text{eq}}$\ and $T_{\text{eq}}$, thus causing any
additive function of $t$\ that may appear in (\ref{c23.5}) to be zero. \ (Note
that at $t=0$, (\ref{c23.5})\ satisfies (\ref{c23.04}) for the problem
discussed above.) \ As we can see, the two diffusion equations (\ref{c23.2})
and (\ref{c23.4}) are consistent with constraint (\ref{c23.5}). \ Furthermore,
they imply that in the absence of mechanical forces, a non-equilibrium
problem, such as the one described in the previous paragraph, equilibrates by
pure diffusion. \ Equation (\ref{c23.2}) is Fick's law which describes
diffusion driven by gradients in mass density.

On the other hand, if we substitute the zero velocity solution (\ref{c23.1})
into the linearized NSF equations (\ref{c16.3})-(\ref{c16.5}), then we find
only the trivial solution,%
\[
\delta\!m\left(  x,t\right)  =\delta\!T\left(  x,t\right)  =0,
\]
unless we examine the special case, $\left(  \alpha_{P}\right)  _{\text{eq}}=0
$,\ which yields%
\[
\delta\!m\left(  x,t\right)  =0
\]
and%
\[
\frac{\partial\delta\!T}{\partial t}=\left(  \frac{k_{F}}{mc_{V}}\right)
_{\text{eq}}\frac{\partial^{2}\delta\!T}{\partial x^{2}}.
\]
Either way, it is clear that for the initial value problem described
previously, the mass density does not equilibrate via pure diffusion when the
NSF formulation is used. \ In fact, a non-zero velocity solution arises.

The results discussed in this section will be demonstrated explicitly in a
future paper \cite{M1d} in which I utilize Green's functions to solve the
problem described here for the general $\overline{M}$-formulation.

\section{\label{sound}Sound Propagation}

Next, let us employ linearized $\overline{M}\left(  D,\eta\right)  $-system
(\ref{c7})-(\ref{c9})\ to study one-dimensional Cartesian sound propagation
for low amplitude sound waves of angular frequency $\omega$. \ For this
problem, one postulates a solution proportional to%
\[
\exp\left(  \kappa x+i\omega t\right)  ,
\]
where $\omega$\ is real and $\kappa$\ is complex, leading to the following
dispersion relation:%
\begin{equation}
\left[
\begin{array}
[c]{c}%
iD_{\text{eq}}^{3}\kappa^{6}+D_{\text{eq}}\left(  -ic_{\text{eq}}%
^{2}+3D_{\text{eq}}\omega\right)  \kappa^{4}+\\
\left(  -3iD_{\text{eq}}\omega-c_{\text{eq}}^{2}\right)  \omega\kappa
^{2}-\omega^{3}%
\end{array}
\right]  =0.\label{c24}%
\end{equation}
Equation (\ref{c24}) may be solved to obtain six $\kappa$-roots:\ the
propagational pair,%
\begin{equation}
\kappa_{p\left(  \pm\right)  }\left(  \omega\right)  \approx\pm\frac{\omega
}{c_{\text{eq}}}\left[  i+\left(  \frac{D}{c^{2}}\right)  _{\text{eq}}%
\omega\right]  ,\label{c25}%
\end{equation}
the thermal pair,\footnote{As discussed in Morse and Ingard \cite[\S 6.4]%
{MandI}, the thermal roots are used to model thermal boundary layers that may
form near walls--see \S \ref{soundbc}.}%
\begin{equation}
\kappa_{t\left(  \pm\right)  }\left(  \omega\right)  =\pm\sqrt{\frac{\omega
}{2D_{\text{eq}}}}\left(  1+i\right)  ,\label{c26}%
\end{equation}
and what I have termed the source pair,\footnote{This is because they
correspond to boundary layers that form next to vibrating sound sources.}%
\begin{equation}
\kappa_{s\left(  \pm\right)  }\left(  \omega\right)  \approx\pm\frac{\omega
}{c_{\text{eq}}}\left[  i+\left(  \frac{c^{2}}{D}\right)  _{\text{eq}}\frac
{1}{\omega}\right]  .\label{c27}%
\end{equation}
The thermal roots (\ref{c26}) are exact solutions to (\ref{c24}), but
(\ref{c25}) and (\ref{c27}) represent approximate solutions in the
hydrodynamic regime, requiring%
\begin{equation}
\left(  \frac{D}{c^{2}}\right)  _{\text{eq}}\omega\ll1.\label{c28}%
\end{equation}
Note that (\ref{c25}) suggests sound attenuation experiments, such as
Greenspan's \cite{green56} and \cite{green59}, may be used to measure the
diffusion parameter $D$\ of the $\overline{M}\left(  D,\eta\right)
$-formulation for various fluids under a wide variety of thermodynamic
conditions. \ Commonly, one finds $\alpha/f^{2}$ and $c$\ data tabulated from
such experiments at given temperatures and pressures, where%
\begin{equation}
\alpha\equiv\operatorname{Re}\kappa_{p\left(  +\right)  }\text{ and }%
f\equiv\frac{\omega}{2\pi},\label{c28.5}%
\end{equation}
and $\alpha/f^{2}$\ may be considered a frequency-independent quantity
provided the acoustic frequency $f$\ is much higher than the relaxation
frequencies for diatomic or polyatomic molecules that may be present. \ In the
hydrodynamic regime, equation (\ref{c25}) implies the formula,%
\begin{equation}
D_{\text{eq}}=\frac{c_{\text{eq}}^{3}}{\left(  2\pi\right)  ^{2}}\frac{\alpha
}{f^{2}},\label{c28.6}%
\end{equation}
and in \textsc{appendix \ref{difp}}, this formula is used together with
acoustical data to compute the diffusion parameter\ for a variety of selected
gases and liquids.

For the NSF formulation, one may use the foregoing procedure to obtain the
dispersion relation,%
\begin{equation}
\left\{
\begin{array}
[c]{c}%
\left[  -i\left(  c^{2}\Sigma\right)  _{\text{eq}}+\left[  E\!u\left(
\frac{\zeta_{NS}}{m}+\frac{4\eta}{3m}\right)  \frac{\eta}{m}\right]
_{\text{eq}}\omega\right]  \kappa^{4}+\\
\left[  -i\left(  2\Gamma+\Sigma\right)  _{\text{eq}}\omega-c_{\text{eq}}%
^{2}\right]  \omega\kappa^{2}-\omega^{3}%
\end{array}
\right\}  =0,\label{c29}%
\end{equation}
where $\Sigma$\ and $\Gamma$\ are defined in equations (\ref{c18.1}) and
(\ref{c18.2}). \ Solving equation (\ref{c29}), one finds four $\kappa
$-roots:\ the propagational pair,%
\begin{equation}
\kappa_{p\left(  \pm\right)  }\left(  \omega\right)  \approx\pm\frac{\omega
}{c_{\text{eq}}}\left[  i+\left(  \frac{\Gamma}{c^{2}}\right)  _{\text{eq}%
}\omega\right]  ,\label{c30}%
\end{equation}
and the thermal pair,%
\begin{equation}
\kappa_{t\left(  \pm\right)  }\left(  \omega\right)  \approx\pm\sqrt
{\frac{\omega}{2\Sigma_{\text{eq}}}}\left\{  \left[  1-\left(  \frac{\Omega
}{c^{2}}\right)  _{\text{eq}}\omega\right]  +i\left[  1+\left(  \frac{\Omega
}{c^{2}}\right)  _{\text{eq}}\omega\right]  \right\}  ,\label{c31}%
\end{equation}
where we have defined%
\begin{equation}
\Omega=\frac{1}{2}\left\{  \left(  1-\frac{1}{\gamma}\right)  \left[
E\!u\frac{\eta}{m}-\gamma\left(  \frac{\zeta_{NS}}{m}+\frac{4\eta}{3m}\right)
\right]  \right\}  ,\label{c31.1}%
\end{equation}
and all four roots are approximations in the hydrodynamic regime,%
\begin{equation}
\left(  \frac{\eta}{mc^{2}}\right)  _{\text{eq}}\omega\ll1.\label{c32}%
\end{equation}

Note that a comparison of equations (\ref{c25}) and (\ref{c30})/(\ref{c18.2})
gives a convenient way of relating the diffusion coefficient, $D_{\text{eq}}%
$,\ of the $\overline{M}\left(  D,\eta\right)  $-formulation to the bulk and
shear viscosities, $\left(  \zeta_{NS}\right)  _{\text{eq}}$\ and
$\eta_{\text{eq}}$, and the thermal conductivity $\left(  k_{F}\right)
_{\text{eq}}$ appearing in the NSF formulation, i.e. to match these
predictions for sound attenuation, one chooses%
\begin{equation}
D_{\text{eq}}=\Gamma_{\text{eq}}=\frac{1}{2}\left\{  \frac{\zeta_{NS}}%
{m}+\left[  \frac{4}{3}+\left(  1-\frac{1}{\gamma}\right)  E\!u\right]
\frac{\eta}{m}\right\}  _{\text{eq}}.\label{c33}%
\end{equation}
For example, in a classical monatomic ideal gas for which one typically
chooses (\ref{e29}) and%
\begin{equation}
\zeta_{NS}=0\text{ and }E\!u=\frac{5}{2},\label{c34}%
\end{equation}
equation (\ref{c33}) implies%
\begin{equation}
D_{\text{eq}}=\frac{7}{6}\left(  \frac{\eta}{m}\right)  _{\text{eq}%
},\label{c35}%
\end{equation}
which in view of equation (I.72), written for this linearized problem as%
\begin{equation}
m_{\text{eq}}D_{\text{eq}}=C_{\text{eq}}\eta_{\text{eq}},\label{c35.5}%
\end{equation}
yields%
\begin{equation}
C_{\text{eq}}=7/6.\label{c36}%
\end{equation}

\begin{figure}
[ptb]
\begin{center}
\includegraphics[
height=3.474in,
width=4.5455in
]%
{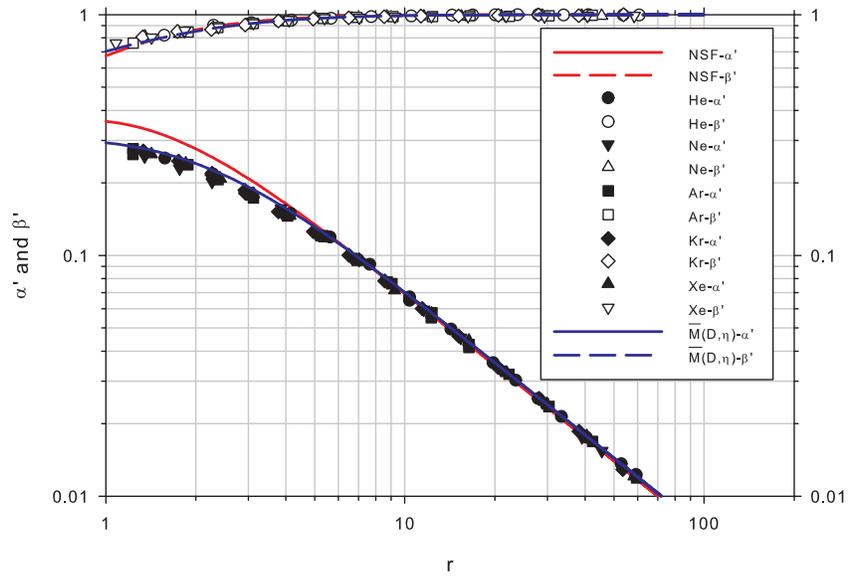}%
\caption{sound propagation in the noble gases}%
\label{fig1}%
\end{center}
\end{figure}

Next, instead of the hydrodynamic regime approximations, (\ref{c25}) and
(\ref{c30}), let us consider the exact propagational roots:%
\begin{equation}
\kappa_{p\left(  +\right)  }\left(  \omega\right)  =\sqrt{\frac{c_{\text{eq}%
}^{2}+2iD_{\text{eq}}\omega-c_{\text{eq}}\sqrt{c_{\text{eq}}^{2}%
+4iD_{\text{eq}}\omega}}{2D_{\text{eq}}^{2}}}\label{c37}%
\end{equation}
corresponding to the $\overline{M}\left(  D,\eta\right)  $-formulation and%
\begin{gather}
\kappa_{p\left(  +\right)  }\left(  \omega\right)  =\nonumber\\
\sqrt{\frac{\left(
\begin{array}
[c]{c}%
c_{\text{eq}}^{2}+i\omega\left[  \left(  \frac{4}{3}+E\!u\right)  \frac{\eta
}{m}+\frac{\zeta_{NS}}{m}\right]  _{\text{eq}}-\\
\sqrt{%
\begin{array}
[c]{c}%
\left\{  c_{\text{eq}}^{2}+i\omega\left[  \left(  \frac{4}{3}+E\!u\right)
\frac{\eta}{m}+\frac{\zeta_{NS}}{m}\right]  _{\text{eq}}\right\}  ^{2}+\\
4\left(  E\!u\frac{\eta}{m}\right)  _{\text{eq}}\omega\left[  -i\left(
\frac{c^{2}}{\gamma}\right)  _{\text{eq}}+\left(  \frac{4\eta}{3m}+\frac
{\zeta_{NS}}{m}\right)  _{\text{eq}}\omega\right]
\end{array}
}%
\end{array}
\right)  }{2\left(  E\!u\frac{\eta}{m}\right)  _{\text{eq}}\left[  -i\left(
\frac{c^{2}}{\gamma}\right)  _{\text{eq}}\frac{1}{\omega}+\left(  \frac{4\eta
}{3m}+\frac{\zeta_{NS}}{m}\right)  _{\text{eq}}\right]  }}\label{c38}%
\end{gather}
corresponding to the NSF formulation. \ \textsc{Figure} \ref{fig1}\ is a plot
of the dimensionless real and imaginary parts,%
\begin{equation}
\alpha^{\prime}=\operatorname{Re}\left(  \frac{c_{\text{eq}}\kappa_{p\left(
+\right)  }}{\omega}\right) \label{c39}%
\end{equation}
and%
\begin{equation}
\beta^{\prime}=\operatorname{Im}\left(  \frac{c_{\text{eq}}\kappa_{p\left(
+\right)  }}{\omega}\right)  ,\label{c40}%
\end{equation}
of the above roots versus the dimensionless parameter,%
\begin{equation}
r\equiv\left(  \frac{mc^{2}}{\gamma\eta}\right)  _{\text{eq}}\frac{1}{\omega
},\label{c41}%
\end{equation}
which is inversely proportional to the Knudsen number for a gas. \ The
$\alpha^{\prime}$\ and $\beta^{\prime}$\ parameters characterize the sound
attenuation and the inverse phase speed, respectively. \ The curves (blue for
the $\overline{M}\left(  D,\eta\right)  $-formulation and red for NSF)
correspond to a classical monatomic ideal gas for which (\ref{e29}),
(\ref{c34}) and (\ref{c35}) are chosen, as well as equation (\ref{e30}) for
the adiabatic sound speed. \ The points are Greenspan's \cite{green56} sound
data for the noble gases, helium, neon, argon, krypton, and xenon, at room
temperature. \ Both axes are logarithmic scale, and the upper and lower curves
correspond to $\beta^{\prime}$ and $\alpha^{\prime}$, respectively. \
From \textsc{figure} \ref{fig1}, it is clear that my theory's sound
predictions match those of the NSF formulation in the hydrodynamic (high $r$,
or low $K\!n$) regime, as intended. \ As I have emphasized throughout, the
$\overline{M}\left(  D,\eta\right)  $-formulation is a continuum theory
appropriate for the $K\!n\ll1$\ regime, and so I make no general claims that
it works well into more rarefied higher Knudsen number regimes. \ However,
\textsc{figure} \ref{fig1} provides an intriguing demonstration that for this
problem, it does significantly better than Navier-Stokes-Fourier there. \ Of
course, until I can provide a formal explanation, this should be viewed as
strictly fortuitous.

In \cite{M1d}, I will use Green's functions to solve the linearized general
$\overline{M}$- formulation in order to fully describe planar sound waves
emanating from a sinusoidally vibrating source into a fluid occupying infinite
space. Afterwards, I will use this solution to interpret the sound roots in a
more precise physical manner and relate them to the quantities that are
measured in sound attenuation experiments such as Greenspan's.

\section{\label{soundbc}Sound at a Boundary}

Morse and Ingard \cite[\textsc{Ch.} 6]{MandI} use the NSF equations in order
to study low amplitude sound waves hitting a wall. \ For easier comparison
with their treatment and for general convenience, as well, let us use the
thermodynamic relationships (\ref{e4}) and (\ref{e5}) to recast the
$\overline{M}\left(  D,\eta\right)  $-formulation (\ref{s1})/(\ref{s2}%
)--assuming variation is in the $x_{1}\equiv x$ direction only, $v_{1}\equiv
v$\ is the only non-zero component of the velocity, and there are no body
forces--in terms of the variables, $\overline{m}$, $P$, $v$, and $T$, and then
linearize about the constant equilibrium state,%
\begin{equation}
\left(  \overline{m},P,v,T\right)  =\left(  m_{\text{eq}},P_{\text{eq}%
},0,T_{\text{eq}}\right)  ,\label{c101n}%
\end{equation}
via%
\begin{align}
\overline{m}  & =m_{\text{eq}}+\delta\!\overline{m}\label{c102}\\
P  & =P_{\text{eq}}+\delta\!P\label{c103}\\
v  & =\delta\!v\label{c104}\\
T  & =T_{\text{eq}}+\delta\!T.\label{c105}%
\end{align}
Assuming%
\begin{align}
\left\vert \delta\!\overline{m}\right\vert  & \ll m_{\text{eq}}\nonumber\\
\left\vert \delta\!P\right\vert  & \ll P_{\text{eq}}\label{c105.5}\\
\left\vert \delta\!T\right\vert  & \ll T_{\text{eq}}\nonumber\\
\left\vert \delta\!v\right\vert  & \ll c_{\text{eq}},\nonumber
\end{align}
one obtains the following approximate system:%
\begin{align}
\frac{\partial\delta\!\overline{m}}{\partial t}  & =-m_{\text{eq}}%
\frac{\partial\delta\!v}{\partial x}\label{c106}\\
\frac{\partial\delta\!P}{\partial t}  & =D_{\text{eq}}\frac{\partial^{2}%
\delta\!P}{\partial x^{2}}-\left(  mc^{2}\right)  _{\text{eq}}\frac
{\partial\delta\!v}{\partial x}\label{c107}\\
\frac{\partial\delta\!v}{\partial t}  & =-\frac{1}{m_{\text{eq}}}%
\frac{\partial\delta\!P}{\partial x}+D_{\text{eq}}\frac{\partial^{2}\delta
\!v}{\partial x^{2}}\label{c108}\\
\frac{\partial\delta\!T}{\partial t}  & =D_{\text{eq}}\frac{\partial^{2}%
\delta\!T}{\partial x^{2}}-\left(  \frac{T\alpha_{P}c^{2}}{c_{P}}\right)
_{\text{eq}}\frac{\partial\delta\!v}{\partial x},\label{c109}%
\end{align}
where $c_{P}$\ is the isobaric specific heat. \ Again, we see that the
mechanical mass equation (\ref{c106}) may be uncoupled from the rest of the
system, and postulating a solution%
\[
\left[
\begin{array}
[c]{c}%
\delta\!P\\
\delta\!v\\
\delta\!T
\end{array}
\right]
\]
to the remaining system (\ref{c107})-(\ref{c109}) that is proportional to%
\[
\exp\left(  \kappa x+i\omega t\right)
\]
with $\omega$ real and $\kappa$ complex, unsurprisingly leads to the same
dispersion relation (\ref{c24}) and the same six roots (\ref{c25})-(\ref{c27})
as before. \ However, when my equations are cast in the present form, one
notes the following interesting fact: the temperature equation (\ref{c109})
may be uncoupled from the pressure and velocity equations, so that the
dispersion relation arising from the reduced system, (\ref{c107}) and
(\ref{c108}), yields only the propagational roots (\ref{c25}) and the source
roots (\ref{c27}). \ The thermal roots (\ref{c26}) arise only when the
temperature equation (\ref{c109}) is coupled back into the system. \ This
means that within the $\overline{M}\left(  D,\eta\right)  $-formulation,
thermal modes of the solution may contribute only to the temperature variable
and not the pressure or velocity variables, i.e. the mechanical variables.
\ On the other hand, propagational and source modes may contribute to all three.

Next, let us examine the length scales associated with each of these three
types of roots. \ If we compute the length scale as the distance it takes for
a mode to attenuate to $1/e$ of its amplitude, then we find the propagational,
thermal, and source mode lengths to be%
\begin{equation}
l_{p}\equiv\frac{1}{\left\vert \operatorname{Re}\kappa_{p\left(  \pm\right)
}\right\vert }=\frac{1}{4\pi^{2}f^{2}}\left(  \frac{c^{3}}{D}\right)
_{\text{eq}},\label{c111n}%
\end{equation}%
\begin{equation}
l_{t}\equiv\frac{1}{\left\vert \operatorname{Re}\kappa_{t\left(  \pm\right)
}\right\vert }=\sqrt{\frac{D_{\text{eq}}}{\pi f}},\label{c112n}%
\end{equation}
and%
\begin{equation}
l_{s}\equiv\frac{1}{\left\vert \operatorname{Re}\kappa_{s\left(  \pm\right)
}\right\vert }=\left(  \frac{D}{c}\right)  _{\text{eq}},\label{c113n}%
\end{equation}
respectively, where (\ref{c25})-(\ref{c27}) and (\ref{c28.5}) have been used
in the above. \ For example, if we consider ultrasonic waves at frequency
$f=11$ MHz as in Greenspan \cite{green56} and argon gas at normal temperature
and pressure ($T_{\text{eq}}=300$ K and $P_{\text{eq}}=1.013\times10^{5}$ Pa)
with $D_{\text{eq}}=1.64\times10^{-5}$ m$^{2}$/s$\ $(see \textsc{appendix
\ref{difp}}) and $c_{\text{eq}}=323$ m/s from the classical monatomic ideal
gas formula (\ref{e30}), then the above length scales are computed to be%
\[
l_{p}=4.30\times10^{-4}\text{ m,}%
\]%
\[
l_{t}=6.87\times10^{-7}\text{ m,}%
\]
and%
\[
l_{s}=5.08\times10^{-8}\text{ m.}%
\]
In this case--and in general, for all ideal gases in the hydrodynamic
regime--the source mode has a length scale $l_{s}$ roughly equal to the mean
free path length of the gas in its equilibrium state $\left(  T_{\text{eq}%
},P_{\text{eq}}\right)  $, regardless of sound frequency, and%
\begin{equation}
l_{p}\gg l_{t}>l_{s}.\label{c114n}%
\end{equation}
Away from boundaries, the propagational modes of sound waves dominate.
\ However, depending on the boundary conditions, the thermal and source modes
may play an important role near walls, causing the formation of boundary
layers with lengths $l_{t}$\ and $l_{s}$, respectively.

When the NSF equations (\ref{s3})/(\ref{s4}) are recast using (\ref{e4}) and
(\ref{e5}) and linearized as above, we arrive at the following system:%
\begin{align}
\frac{\partial\delta\!P}{\partial t}  & =\left(  \frac{\alpha_{P}c^{2}k_{F}%
}{c_{P}}\right)  _{\text{eq}}\frac{\partial^{2}\delta\!T}{\partial x^{2}%
}-\left(  mc^{2}\right)  _{\text{eq}}\frac{\partial\delta\!v}{\partial
x}\label{c120n}\\
\frac{\partial\delta\!v}{\partial t}  & =-\frac{1}{m_{\text{eq}}}%
\frac{\partial\delta\!P}{\partial x}+\left(  \frac{\zeta_{NS}}{m}+\frac{4}%
{3}\frac{\eta}{m}\right)  _{\text{eq}}\frac{\partial^{2}\delta\!v}{\partial
x^{2}}\label{c121n}\\
\frac{\partial\delta\!T}{\partial t}  & =\left(  \frac{\gamma k_{F}}{mc_{P}%
}\right)  _{\text{eq}}\frac{\partial^{2}\delta\!T}{\partial x^{2}}-\left(
\frac{T\alpha_{P}c^{2}}{c_{P}}\right)  _{\text{eq}}\frac{\partial\delta
\!v}{\partial x},\label{c122n}%
\end{align}
which again yields dispersion relation (\ref{c29}) and the four roots
(\ref{c30}) and (\ref{c31}). \ Unlike in the $\overline{M}\left(
D,\eta\right)  $-formulation, none of the above equations uncouple from one
another. \ Therefore, in the NSF formulation the two possible modes,
propagational and thermal, may contribute to each of the three variables:
pressure, velocity, and temperature.

Using (\ref{c30}) and (\ref{c31}), the length scales associated with the
propagational and thermal modes of the NSF formulation are computed to be%
\begin{equation}
l_{p}\equiv\frac{1}{\left\vert \operatorname{Re}\kappa_{p\left(  \pm\right)
}\right\vert }=\frac{c_{\text{eq}}^{3}}{4\pi^{2}f^{2}\Gamma_{\text{eq}}%
}\label{c123n}%
\end{equation}
and%
\begin{equation}
l_{t}\equiv\frac{1}{\left\vert \operatorname{Re}\kappa_{t\left(  \pm\right)
}\right\vert }=\sqrt{\frac{\Sigma_{\text{eq}}}{\pi f}}\frac{1}{1-2\pi f\left(
\frac{\Omega}{c^{2}}\right)  _{\text{eq}}}.\label{c124n}%
\end{equation}
In view of (\ref{c33}), equation (\ref{c123n}) yields the same propagation
length as (\ref{c111n}). \ Also, in the hydrodynamic regime, the thermal
length given by (\ref{c124n}) is the same order of magnitude as the thermal
length (\ref{c112n}) from my formulation.

Let us assume the $yz$-plane forms a sound barrier at $x=0$\ for sound waves
coming from the $+x$-direction and, as in Morse and Ingard, define the
acoustic impedance of the wall as%
\begin{equation}
z_{w}\equiv\frac{\text{acoustic pressure at wall}}{\text{normal fluid velocity
at wall}}=-\left.  \frac{\delta\!P}{\delta\!v}\right\vert _{x=0}.\label{c115n}%
\end{equation}
Infinite impedance corresponds to a perfectly reflected sound wave in which
its outgoing velocity is equal in magnitude and opposite in direction to its
incoming velocity, resulting in%
\begin{equation}
\left.  \delta\!v\right\vert _{x=0}=0.\label{c116n}%
\end{equation}
For non-infinite wall impedances, however, there is a non-zero normal fluid
velocity at the wall. \ Using the $\overline{M}\left(  D,\eta\right)
$-formulation, we may still consider the wall to be both stationary and
impermeable by enforcing the no total mass flux condition,%
\begin{equation}
\left.  \underline{j}_{M}\cdot\underline{n}\right\vert _{x=0}=0\label{c116.3n}%
\end{equation}
where, in this case, the normal vector $\underline{n}$ is the unit vector in
the $-x$ direction. \ With (\ref{s7})/(\ref{s2} a) linearized and
thermodynamic relation (\ref{e5}), the above condition implies%
\begin{equation}
\left.  \left\{  -D_{\text{eq}}\left[
\begin{array}
[c]{c}%
-\left(  m\alpha_{P}\right)  _{\text{eq}}\frac{\delta\!T}{dx}+\\
\left(  \frac{\gamma}{c^{2}}\right)  _{\text{eq}}\frac{\delta\!P}{dx}%
\end{array}
\right]  +m_{\text{eq}}\delta\!v\right\}  \right\vert _{x=0}=0.\label{c118n}%
\end{equation}
On the other hand, if (\ref{c116n}) is not satisfied in the NSF formulation,
there arises the curious situation of a non-zero normal mass flux at $x=0$.
\ Morse and Ingard \cite[p. 260]{MandI} explain this phenomenon as
follows:\ "The acoustic pressure acts on the surface and tends to make it
move, or else tends to force more fluid into the pores of the surface." \ Wall
movement and/or fluid lost to pores in the wall are yet never treated explicitly.

I will make the above ideas precise in a future paper \cite{M1dSID} in which
Green's functions on a semi-infinite domain are used to solve linearized
one-dimensional problems involving fluid disturbances interacting with an
impedance wall.

\section{\label{flh}Hydrodynamical Fluctuations}

The procedure for investigating stationary-Gaussian-Markov processes, as
described by Fox and Uhlenbeck \cite{fox}, may be used to derive the formulas
of hydrodynamical fluctuations for the $\overline{M}\left(  D,\eta\right)
$-formulation.\footnote{\label{fnhf}I carry out a complete analysis of
linearized hydrodynamical fluctuations for the general $\overline{M}%
$-formulation in a future paper \cite{stochIII}, in which I derive in detail
the formulas contained in this section and the following section on light
scattering.} \ This involves using the linearized three-dimensional Cartesian
version of the hydrodynamic equations (I.59) with transport coefficients given
by (I.67), (I.68), and (I.61 b), interpreting the variables as stochastically
fluctuating variables, and introducing zero-mean fluctuating hydrodynamic
forces. \ This yields the following Langevin type of system:%
\begin{align}
\frac{\partial\delta\!m}{\partial t}  & =D_{\text{eq}}\nabla^{2}%
\delta\!m-\nabla\cdot\delta\!\underline{p}-\nabla\cdot\underline{J}%
_{M}\nonumber\\
\frac{\partial\delta\!\underline{p}}{\partial t}  & =\left[
\begin{array}
[c]{c}%
\left(  \beta_{m}\right)  _{\text{eq}}\nabla\delta\!m+\left(  \beta
_{u}\right)  _{\text{eq}}\nabla\delta\!u+\\
\left(  \frac{\eta}{m}\right)  _{\text{eq}}\nabla^{2}\delta\!\underline{p}+\\
\left(  D-\frac{\eta}{m}\right)  _{\text{eq}}\nabla\left(  \nabla\cdot
\delta\!\underline{p}\right)
\end{array}
\right]  -\nabla\cdot\underline{\underline{J}}_{\underline{P}}\label{c43}\\
\frac{\partial\delta\!u}{\partial t}  & =D_{\text{eq}}\nabla^{2}%
\delta\!u-\left(  h_{M}\right)  _{\text{eq}}\nabla\cdot\delta\!\underline
{p}-\nabla\cdot\underline{J}_{U}.\nonumber
\end{align}
In the above, the mechanical mass density equation has been uncoupled. \ Also,
$\underline{J}_{M}$, $\underline{\underline{J}}_{\underline{P}}$, and
$\underline{J}_{U}$ represent the fluctuating mass, momentum, and internal
energy fluxes, respectively, with $\underline{\underline{J}}_{\underline{P}}$
assumed to be symmetric and%
\begin{equation}
\left\langle \underline{J}_{M}\right\rangle =\left\langle \underline{J}%
_{U}\right\rangle =\underline{0}\text{ and }\left\langle \underline
{\underline{J}}_{\underline{P}}\right\rangle =\underline{\underline{0}%
},\label{c44}%
\end{equation}
where $\left\langle \_\right\rangle $\ denotes a stochastic average.

Following the steps outlined in Fox and Uhlenbeck leads to a generalized
fluctuation-dissipation theorem for stationary-Gaussian-Markov processes
which, when expressed for the problem at hand, yields the following
correlation formulas for the Cartesian components of the fluctuating fluxes:%
\begin{gather}
\left\langle \left(  J_{M}\right)  _{i}\left(  \underline{x},t\right)  \left(
J_{M}\right)  _{j}\left(  \underline{x}^{\prime},t^{\prime}\right)
\right\rangle =\nonumber\\
2V\left\langle \delta\!m^{2}\right\rangle _{V,\text{eq}}D_{\text{eq}}%
\delta\left(  \underline{x}-\underline{x}^{\prime}\right)  \delta\left(
t-t^{\prime}\right)  \delta_{ij},\label{c45}%
\end{gather}%
\begin{gather}
\left\langle \left(  J_{U}\right)  _{i}\left(  \underline{x},t\right)  \left(
J_{U}\right)  _{j}\left(  \underline{x}^{\prime},t^{\prime}\right)
\right\rangle =\nonumber\\
2V\left\langle \delta\!u^{2}\right\rangle _{V,\text{eq}}D_{\text{eq}}%
\delta\left(  \underline{x}-\underline{x}^{\prime}\right)  \delta\left(
t-t^{\prime}\right)  \delta_{ij},\label{c46}%
\end{gather}%
\begin{gather}
\left\langle \left(  J_{M}\right)  _{i}\left(  \underline{x},t\right)  \left(
J_{U}\right)  _{j}\left(  \underline{x}^{\prime},t^{\prime}\right)
\right\rangle =\nonumber\\
2V\left\langle \delta\!m\delta\!u\right\rangle _{V,\text{eq}}D_{\text{eq}%
}\delta\left(  \underline{x}-\underline{x}^{\prime}\right)  \delta\left(
t-t^{\prime}\right)  \delta_{ij},\label{c50}%
\end{gather}%
\begin{gather}
\left\langle \left(  J_{\underline{P}}\right)  _{ij}\left(  \underline
{x},t\right)  \left(  J_{\underline{P}}\right)  _{kl}\left(  \underline
{x}^{\prime},t^{\prime}\right)  \right\rangle =\nonumber\\
\left\{
\begin{array}
[c]{l}%
2k_{B}\left(  mTD\right)  _{\text{eq}}\delta\left(  \underline{x}%
-\underline{x}^{\prime}\right)  \delta\left(  t-t^{\prime}\right)  \text{ if
}i=j=k=l\\
2k_{B}\left(  T\eta\right)  _{\text{eq}}\delta\left(  \underline{x}%
-\underline{x}^{\prime}\right)  \delta\left(  t-t^{\prime}\right)  \text{ if
}i=k\text{, }j=l\text{, }i\neq j\\
2k_{B}\left(  T\eta\right)  _{\text{eq}}\delta\left(  \underline{x}%
-\underline{x}^{\prime}\right)  \delta\left(  t-t^{\prime}\right)  \text{ if
}i=l\text{, }j=k\text{, }i\neq j\\
2k_{B}\left[  T\left(  mD-2\eta\right)  \right]  _{\text{eq}}\delta\left(
\underline{x}-\underline{x}^{\prime}\right)  \delta\left(  t-t^{\prime
}\right)  \text{ if }\left(
\begin{array}
[c]{c}%
i=j\text{,}\\
k=l\text{,}\\
i\neq k
\end{array}
\right) \\
0\text{ otherwise}%
\end{array}
\right.  ,\label{c47}%
\end{gather}
and%
\begin{equation}
\left\langle \left(  J_{\underline{P}}\right)  _{ij}\left(  \underline
{x},t\right)  \left(  J_{M}\right)  _{k}\left(  \underline{x}^{\prime
},t^{\prime}\right)  \right\rangle =\left\langle \left(  J_{\underline{P}%
}\right)  _{ij}\left(  \underline{x},t\right)  \left(  J_{U}\right)
_{k}\left(  \underline{x}^{\prime},t^{\prime}\right)  \right\rangle
=0.\label{c48}%
\end{equation}
In the above formulas, the indices, $i$, $j$, and $k$, may equal $1$, $2$, or
$3$ (the three Cartesian directions), $\delta$ is used to represent Dirac
delta distributions, $\delta_{ij}$ is the Kronecker delta, $k_{B}$\ denotes
the Boltzmann constant, and I have employed equilibrium thermodynamic
fluctuation formulas (\ref{e40})-(\ref{e42}), which apply to a system of fixed
volume $V$ in contact with a heat/particle reservoir. \ Note the similarity in
form of equations (\ref{c45})-(\ref{c50}).

For comparison, the Langevin system corresponding to the NSF formulation is%
\begin{align}
\frac{\partial\delta\!m}{\partial t}  & =-\nabla\cdot\delta\!\underline
{p}-\nabla\cdot\underline{J}_{M}\nonumber\\
\frac{\partial\delta\!\underline{p}}{\partial t}  & =\left\{
\begin{array}
[c]{c}%
\left(  \beta_{m}\right)  _{\text{eq}}\nabla\delta\!m+\left(  \beta
_{u}\right)  _{\text{eq}}\nabla\delta\!u+\\
\left(  \frac{\eta}{m}\right)  _{\text{eq}}\nabla^{2}\delta\!\underline{p}+\\
\left(  \frac{\zeta_{NS}+\eta/3}{m}\right)  _{\text{eq}}\nabla\left(
\nabla\cdot\delta\!\underline{p}\right)
\end{array}
\right\}  -\nabla\cdot\underline{\underline{J}}_{\underline{P}}\label{c54}\\
\frac{\partial\delta\!u}{\partial t}  & =\left\{
\begin{array}
[c]{c}%
\left(  \frac{k_{F}}{mc_{V}}\right)  _{\text{eq}}\nabla^{2}\delta\!u-\\
\left[  \frac{\left(  h_{M}-\frac{T\alpha_{P}}{m\kappa_{T}}\right)  k_{F}%
}{mc_{V}}\right]  _{\text{eq}}\nabla^{2}\delta\!m-\\
\left(  h_{M}\right)  _{\text{eq}}\nabla\cdot\delta\!\underline{p}%
\end{array}
\right\}  -\nabla\cdot\underline{J}_{U}\nonumber
\end{align}
from which one may derive Fox and Uhlenbeck's correlations,%
\begin{equation}
\left\langle \left(  J_{M}\right)  _{i}\left(  \underline{x},t\right)  \left(
J_{M}\right)  _{j}\left(  \underline{x}^{\prime},t^{\prime}\right)
\right\rangle =0,\label{c55}%
\end{equation}%
\begin{equation}
\left\langle \left(  J_{U}\right)  _{i}\left(  \underline{x},t\right)  \left(
J_{U}\right)  _{j}\left(  \underline{x}^{\prime},t^{\prime}\right)
\right\rangle =2k_{B}\left(  T^{2}k_{F}\right)  _{\text{eq}}\delta\left(
\underline{x}-\underline{x}^{\prime}\right)  \delta\left(  t-t^{\prime
}\right)  \delta_{ij},\label{c56}%
\end{equation}%
\begin{equation}
\left\langle \left(  J_{M}\right)  _{i}\left(  \underline{x},t\right)  \left(
J_{U}\right)  _{j}\left(  \underline{x}^{\prime},t^{\prime}\right)
\right\rangle =0,\label{c60}%
\end{equation}%
\begin{gather}
\left\langle \left(  J_{\underline{P}}\right)  _{ij}\left(  \underline
{x},t\right)  \left(  J_{\underline{P}}\right)  _{kl}\left(  \underline
{x}^{\prime},t^{\prime}\right)  \right\rangle =\nonumber\\
\left\{
\begin{array}
[c]{l}%
2k_{B}\left[  T\left(  \frac{4\eta}{3}+\zeta_{NS}\right)  \right]
_{\text{eq}}\delta\left(  \underline{x}-\underline{x}^{\prime}\right)
\delta\left(  t-t^{\prime}\right)  \text{ if }i=j=k=l\\
2k_{B}\left(  T\eta\right)  _{\text{eq}}\delta\left(  \underline{x}%
-\underline{x}^{\prime}\right)  \delta\left(  t-t^{\prime}\right)  \text{ if
}i=k\text{, }j=l\text{, }i\neq j\\
2k_{B}\left(  T\eta\right)  _{\text{eq}}\delta\left(  \underline{x}%
-\underline{x}^{\prime}\right)  \delta\left(  t-t^{\prime}\right)  \text{ if
}i=l\text{, }j=k\text{, }i\neq j\\
2k_{B}\left[  T\left(  \zeta_{NS}-\frac{2\eta}{3}\right)  \right]
_{\text{eq}}\delta\left(  \underline{x}-\underline{x}^{\prime}\right)
\delta\left(  t-t^{\prime}\right)  \text{ if }\left(
\begin{array}
[c]{c}%
i=j\text{,}\\
k=l\text{,}\\
i\neq k
\end{array}
\right) \\
0\text{ otherwise}%
\end{array}
\right.  ,\label{c57}%
\end{gather}
and%
\begin{equation}
\left\langle \left(  J_{\underline{P}}\right)  _{ij}\left(  \underline
{x},t\right)  \left(  J_{M}\right)  _{k}\left(  \underline{x}^{\prime
},t^{\prime}\right)  \right\rangle =\left\langle \left(  J_{\underline{P}%
}\right)  _{ij}\left(  \underline{x},t\right)  \left(  J_{U}\right)
_{k}\left(  \underline{x}^{\prime},t^{\prime}\right)  \right\rangle
=0.\label{c58}%
\end{equation}
Therefore, in the NSF formulation,%
\begin{equation}
\underline{J}_{M}=\underline{0},\label{c63}%
\end{equation}
whereas my theory predicts a non-zero fluctuating mass flux.

This exercise illuminates certain ideas about the thermodynamic nature of a
continuum mechanical point, i.e. a point occupied by matter whose motion is
tracked by its local velocity, $\underline{v}$, and whose infinitesimal volume
is determined by its continuum mechanical deformation. \ As per the local
equilibrium hypothesis, near-equilibrium continuum theories are constructed
under the assumption that any given continuum mechanical point may be viewed
as an equilibrium thermodynamic subsystem in contact with the rest of the
material which acts as a reservoir. \ The question as to whether or not the
fluctuating mass flux $\underline{J}_{M}$ is zero, decides the very nature of
this reservoir. \ All continuum theories view the surrounding material as a
reservoir for heat (so that the point exchanges energy with the surrounding
fluid to maintain a temperature determined by thermodynamics), but is it also
a particle reservoir (so that it exchanges mass with the surrounding fluid to
maintain a thermodynamically determined chemical potential)? \ My theory, with
its non-zero $\underline{J}_{M}$, asserts that it is, whereas the
Navier-Stokes-Fourier theory, in view of its (\ref{c63}) prediction, argues
that it is not, meaning that each continuum mechanical point should be
conceived of as having an impermeable wall surrounding it. \ Although
subsystems and reservoirs are abstract constructs in this setting, I believe
that it makes sense to allow the mass of a continuum mechanical point to
fluctuate. \ This view is reinforced by the natural appearance of fluctuation
formulas (\ref{e40})-(\ref{e42}) for a heat/particle reservoir in my theory's
correlations, (\ref{c45})-(\ref{c50}).

\section{\label{ls}Light Scattering}

One may derive hydrodynamic light scattering spectra for the $\overline
{M}\left(  D,\eta\right)  $- formulation by using the same procedure as that
used in Berne and Pecora \cite[\S 10.4]{berne} for the NSF
formulation.\footnote{See footnote \ref{fnhf}.} \ The main steps involved are
to (1) linearize the Cartesian three-dimensional equations--expressed in terms
of the variables, mechanical mass $\overline{m}$, particle number density
$n_{p}=N_{A}m/A$ where $N_{A}$\ and $A$ respectively denote Avogadro's number
and the atomic weight, divergence of the velocity $\phi=\nabla\cdot
\underline{v}$, and temperature $T$--about the constant equilibrium state,%
\begin{equation}
\left(  \overline{m},n_{p},\phi,T\right)  =\left(  m_{\text{eq}},\left(
n_{p}\right)  _{\text{eq}},0,T_{\text{eq}}\right)  ,\label{c64}%
\end{equation}
(2) uncouple the mechanical mass density equation, (3) take the
Fourier-Laplace transform of the remaining linear system, (4) solve for the
Fourier-Laplace transformed variables, and (5) use this solution to construct
time-correlation functions and their spectral densities. \ Of particular
interest is the particle-particle spectrum, computed for my formulation to be
exactly%
\begin{equation}
S_{N_{p}N_{p}}\left(  k,\omega\right)  =\frac{\mathcal{S}}{\pi}\left(
\begin{array}
[c]{c}%
\left(  1-\frac{1}{\gamma_{\text{eq}}}\right)  \frac{D_{\text{eq}}k^{2}%
}{\omega^{2}+\left(  D_{\text{eq}}k^{2}\right)  ^{2}}+\\
\frac{1}{2\gamma_{\text{eq}}}\left\{
\begin{array}
[c]{c}%
\frac{D_{\text{eq}}k^{2}}{\left[  \omega+\omega_{B}\left(  k\right)  \right]
^{2}+\left(  D_{\text{eq}}k^{2}\right)  ^{2}}+\\
\frac{D_{\text{eq}}k^{2}}{\left[  \omega-\omega_{B}\left(  k\right)  \right]
^{2}+\left(  D_{\text{eq}}k^{2}\right)  ^{2}}%
\end{array}
\right\}
\end{array}
\right)  .\label{c65}%
\end{equation}
In the above, $k$\ is used to represent the magnitude of the wave vector,
$\underline{k}$; $\omega$\ is the angular frequency; $\mathcal{S}$\ is the
structure factor defined as%
\begin{align}
\mathcal{S}  & =\left\langle \delta\!N_{p}^{2}\right\rangle _{V,\text{eq}%
}\label{c66}\\
& =Vk_{B}\left(  n_{p}^{2}T\kappa_{T}\right)  _{\text{eq}}\label{c66.1}%
\end{align}
by equation (\ref{e42.1}) in \textsc{appendix \ref{appetr}} where
$V$\ represents the scattering volume; and $\omega_{B}\left(  k\right)  $\ is
the frequency shift of the Brillouin doublets,%
\begin{equation}
\omega_{B}\left(  k\right)  =c_{\text{eq}}k.\label{c67}%
\end{equation}
The right-hand side of (\ref{c65}) is the sum of three Lorentzian line shapes,
the first of which is the central Rayleigh peak, and the other two, the
negatively and positively $\omega_{B}\left(  k\right)  $-shifted Brillouin
peaks. \ The Brillouin peaks represent light scattering due to the sound
modes, or phonons, generated by pressure fluctuations, and the Rayleigh peak
is caused by light being scattered from specific entropy fluctuations.

The NSF particle-particle spectrum is given by the more complicated
expression,%
\begin{gather}
S_{N_{p}N_{p}}\left(  k,\omega\right)  =\nonumber\\
\frac{\mathcal{S}}{\pi}\operatorname{Re}\left(  \frac{\left\{
\begin{array}
[c]{c}%
-\omega^{2}+i\omega\left[  \frac{\zeta_{NS}}{m}+\left(  \frac{4}%
{3}+E\!u\right)  \frac{\eta}{m}\right]  _{\text{eq}}k^{2}+\\
\left(  1-\frac{1}{\gamma_{\text{eq}}}\right)  \omega_{B}^{2}\left(  k\right)
+\\
\left[  E\!u\left(  \frac{\zeta_{NS}}{m}+\frac{4}{3}\frac{\eta}{m}\right)
\frac{\eta}{m}\right]  _{\text{eq}}k^{4}%
\end{array}
\right\}  }{\left\{
\begin{array}
[c]{c}%
-i\omega^{3}-\omega^{2}\left[  \frac{\zeta_{NS}}{m}+\left(  \frac{4}%
{3}+E\!u\right)  \frac{\eta}{m}\right]  _{\text{eq}}k^{2}+\\
i\omega\left\{  \omega_{B}^{2}\left(  k\right)  +\left[  E\!u\left(
\frac{\zeta_{NS}}{m}+\frac{4}{3}\frac{\eta}{m}\right)  \frac{\eta}{m}\right]
_{\text{eq}}k^{4}\right\}  +\\
\left(  \frac{E\!u}{\gamma}\frac{\eta}{m}\right)  _{\text{eq}}\omega_{B}%
^{2}\left(  k\right)  k^{2}%
\end{array}
\right\}  }\right)  ,\label{c68}%
\end{gather}
where $E\!u$\ is the Euken ratio defined in (\ref{c18}). \ In the hydrodynamic
regime for which%
\begin{equation}
\left(  \frac{\eta}{mc}\right)  _{\text{eq}}k\ll1,\label{c69}%
\end{equation}
equation (\ref{c68}) becomes approximately\footnote{Note that below I have
corrected mistakes appearing in Berne and Pecora's approximate formula
\cite[p. 243 eq. (10.4.38-39)]{berne}.}%
\begin{gather}
S_{N_{p}N_{p}}\left(  k,\omega\right)  =\nonumber\\
\frac{\mathcal{S}}{\pi}\left(
\begin{array}
[c]{c}%
\left(  1-\frac{1}{\gamma_{\text{eq}}}\right)  \frac{\Sigma_{\text{eq}}k^{2}%
}{\omega^{2}+\left(  \Sigma_{\text{eq}}k^{2}\right)  ^{2}}+\\
\frac{1}{2\gamma_{\text{eq}}}\left\{  \frac{\Gamma_{\text{eq}}k^{2}}{\left[
\omega+\omega_{B}\left(  k\right)  \right]  ^{2}+\left(  \Gamma_{\text{eq}%
}k^{2}\right)  ^{2}}+\frac{\Gamma_{\text{eq}}k^{2}}{\left[  \omega-\omega
_{B}\left(  k\right)  \right]  ^{2}+\left(  \Gamma_{\text{eq}}k^{2}\right)
^{2}}\right\}  +\\
\frac{b\left(  k\right)  }{2}\left\{  \frac{\left[  \omega+\omega\left(
k\right)  \right]  }{\left[  \omega+\omega_{B}\left(  k\right)  \right]
^{2}+\left(  \Gamma_{\text{eq}}k^{2}\right)  ^{2}}-\frac{\left[  \omega
-\omega\left(  k\right)  \right]  }{\left[  \omega-\omega_{B}\left(  k\right)
\right]  ^{2}+\left(  \Gamma_{\text{eq}}k^{2}\right)  ^{2}}\right\}
\end{array}
\right)  ,\label{c70}%
\end{gather}
where $\Sigma$\ and $\Gamma$\ are defined in (\ref{c18.1}) and (\ref{c18.2})
and%
\begin{equation}
b\left(  k\right)  \equiv\frac{k^{2}}{\gamma_{\text{eq}}\omega_{B}\left(
k\right)  }\left\{  \frac{\zeta_{NS}}{2m}+\left[  \frac{3}{2}\left(
1-\frac{1}{\gamma}\right)  E\!u+\frac{2}{3}\right]  \frac{\eta}{m}\right\}
_{\text{eq}}.\label{c71}%
\end{equation}
In approximate expression (\ref{c70}), the first three terms represent the
Lorentzian Rayleigh and Brillouin peaks and the last two terms give a fairly
small, but not negligible, non-Lorentzian contribution.

\begin{figure}
[ptb]
\begin{center}
\includegraphics[
height=3.6772in,
width=4.5446in
]%
{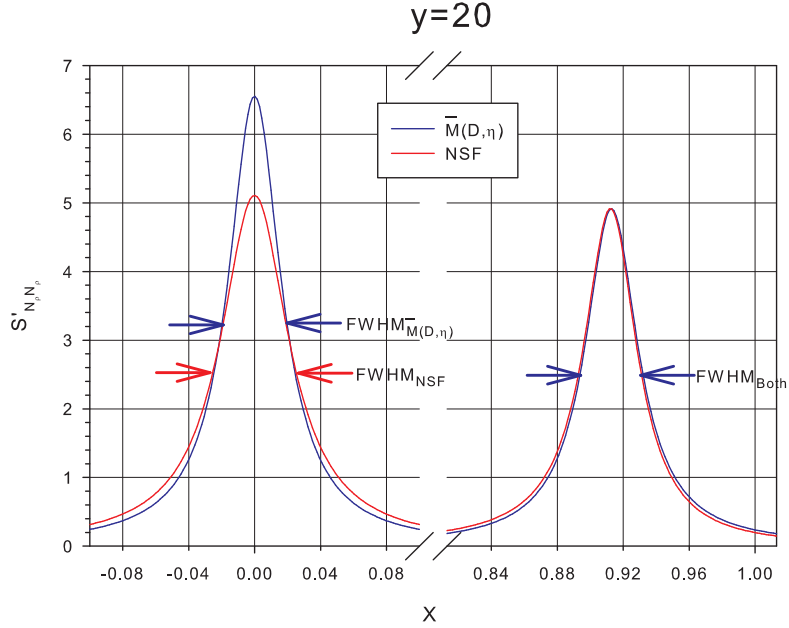}%
\caption{theoretical light scattering spectra for a classical monatomic ideal
gas (y $=20$)}%
\label{fig2}%
\end{center}
\end{figure}

Next, as in Clark \cite{clark}\ and Fookson et al. \cite{fookson}, let us
define\ a dimensionless angular frequency,%
\begin{equation}
\text{x}=\left(  \frac{1}{c}\sqrt{\frac{\gamma}{2}}\right)  _{\text{eq}}%
\frac{\omega}{k},\label{c72}%
\end{equation}
the dimensionless uniformity parameter,%
\begin{equation}
\text{y}=\frac{1}{3k}\left(  \sqrt{\frac{2}{\gamma}}\frac{mc}{\eta}\right)
_{\text{eq}},\label{c73}%
\end{equation}
which is inversely proportional to the Knudsen number for a gas, and the
dimensionless particle-particle spectrum,%
\begin{equation}
S_{N_{p}N_{p}}^{\prime}=\left(  c\sqrt{\frac{2}{\gamma}}\right)  _{\text{eq}%
}\frac{k}{\mathcal{S}}S_{N_{p}N_{p}},\label{c74}%
\end{equation}
which has an area of $1$ when written as a function of x\ and y\ and
integrated over all x. \ \textsc{Figure} \ref{fig2} is a plot of the
dimensionless particle-particle spectra, $S_{N_{p}N_{p}}^{\prime}\left(
\text{x},\text{y}\right)  $, predicted by the $\overline{M}\left(
D,\eta\right)  $-formulation (blue) and the NSF formulation (red) for a
classical monatomic ideal gas with y $=20$ and equations (\ref{c34}),
(\ref{c35}), and (\ref{e30}) assumed. \ The arrows indicate the widths at
half-height. \ (Since the spectra are symmetric about the $0$ frequency, only
the right Brillouin peaks are shown.) \
Both theories predict similar Brillouin peaks,\footnote{This is to be expected
since the Brillouin peaks correspond to the sound modes, and in \S \ref{sound}%
, parameters were chosen such that my formulation and the NSF formulation gave
the same sound predictions in the hydrodynamic regime.} but the $\overline
{M}\left(  D,\eta\right)  $-formulation has a Rayleigh peak with a maximum
about $29\%$ higher and width at half-height about $29\%$ narrower than that
of NSF. \ On the other hand, the areas under both of the Rayleigh peak
predictions are the same, leading to identical Landau-Placzek ratios, i.e.%
\begin{equation}
\frac{\text{area of (central) Rayleigh peak}}{\text{area of both (shifted)
Brillouin peaks}}=\gamma_{\text{eq}}-1\label{c75}%
\end{equation}
which, in view of (\ref{e29}) for a classical monatomic ideal gas, is $2/3$.
\ Note that for the $\overline{M}\left(  D,\eta\right)  $-formulation, the
Brillouin peaks and the Rayleigh peak all have the same width at half-height,
a quantity that is inversely proportional to the lifetime of the excitation
corresponding to the peak. \ This means that my theory predicts the same
lifetime for phonons as it does for excitations caused by specific entropy
fluctuations, whereas the NSF formulation predicts that excitations caused by
specific entropy fluctuations die out sooner than the phonons ($29\%$ sooner
in the case of a classical monatomic ideal gas). \ One other thing to note is
that the Brillouin portion of my exact spectrum (\ref{c65}) predicts phonons
corresponding exactly to the adiabatic sound speed, $c_{\text{eq}}$, whereas
the approximate Navier-Stokes-Fourier spectrum (\ref{c70}) predicts a small
change in speed due to the non-Lorentzian contributions, e.g. in
\textsc{figure} \ref{fig2}, it can be observed that for a classical monatomic
ideal gas, there is a shift of the NSF Brillouin peaks slightly toward the
center frequency.

\section{\label{grav}The Effect of Gravity on the Atmosphere}

Let us limit our investigation to the lowest part of the Earth's atmosphere,
the troposphere. \ Typical discussions of this problem--see Sommerfeld
\cite[\textsc{ch. II} \S 7 pp. 49-55]{som}--do not involve the equations of
fluid dynamics and I briefly summarize the ideas behind this type of treatment
first. \ One begins with the hydrostatic equation for the pressure as a
function of altitude,%
\begin{equation}
P\left(  x\right)  =P_{\text{sl}}-m_{\text{sl}}gx\label{c121}%
\end{equation}
where $x$\ is the altitude above sea level, a subscript of "sl" indicates
values for air at sea level, and $g$\ is the constant acceleration of gravity,
which acts in the $-x$-direction. \ For a classical ideal gas, relationship
(\ref{e25}) allows us to write (\ref{c121}) as%
\begin{equation}
P\left(  x\right)  =P_{\text{sl}}-\frac{A}{R}\left(  \frac{P}{T}\right)
_{\text{sl}}gx.\label{c122}%
\end{equation}
Next, one makes the observation that, in the troposphere, like the pressure,
the temperature also decreases with altitude. \ Furthermore, one obtains this
behavior qualitatively by enforcing the isentropic assumption,%
\begin{equation}
\frac{d\sigma}{dx}=0,\label{c123}%
\end{equation}
where $\sigma$\ represents the specific entropy. \ Using relation (\ref{e28})
for a classical ideal gas with constant specific heat, one finds the
approximation,%
\begin{equation}
\frac{d\sigma}{dx}=\frac{R\gamma}{A\left(  \gamma-1\right)  T_{\text{sl}}%
}\frac{dT}{dx}-\frac{R}{AP_{\text{sl}}}\frac{dP}{dx},\label{c124}%
\end{equation}
holds when $x$ is not too high above sea level, and this relationship, when
used together with (\ref{c122}), implies%
\begin{equation}
\frac{dT}{dx}=-\frac{A\left(  \gamma-1\right)  }{R\gamma}g\label{c125}%
\end{equation}
so that the temperature as a function of altitude is given by%
\begin{equation}
T\left(  x\right)  =T_{\text{sl}}-\frac{A\left(  \gamma-1\right)  }{R\gamma
}gx.\label{c126}%
\end{equation}
Using the values $g=9.81$ m$/$s$^{2}$, $R=8.31$ J/$\left(  \text{mol}%
\cdot\text{K}\right)  $, and the atomic weight and ratio of specific heats for
dry air, $A=0.0290$ kg/m$^{3}$ and $\gamma=1.4$, the quantity $-dT/dx$, called
the temperature lapse rate, is calculated to be%
\begin{equation}
-\frac{dT}{dx}=0.0098\text{ K/m,}\label{c127}%
\end{equation}
which is high when compared to the average measured value presented in
\cite{wtrop}:%
\begin{equation}
-\frac{dT}{dx}=0.0065\text{ K/m,}\label{c128}%
\end{equation}
although these being the same order of magnitude is somewhat of a triumph in
view of all of the simplifying assumptions made in order to obtain (\ref{c127}).

As mentioned earlier, the equations of fluid dynamics are not used at all in
the above arguments. \ Let us also point out that the isentropic assumption
was \textit{chosen} because it gives a temperature law which predicts nature
better than, say, an isothermal assumption does, and it was not, in fact,
derived from other principles. \ Below, I treat this problem using\ the
linearized steady-state equations of fluid dynamics with mass and total energy
balance considerations, and I show that the $\overline{M}\left(
D,\eta\right)  $-formulation predicts the isentropic condition (\ref{c123}),
whereas--under the same assumptions--the NSF formulation predicts an
unphysical isothermal condition.

We once again have a Cartesian one-dimensional problem with variation in the
$x_{1}\equiv x$ direction only, and $v_{1}\equiv v$ as the only non-zero
velocity component. \ For this steady-state problem it is appropriate to use
$\overline{M}\left(  D,\eta\right)  $-formulation (\ref{s10}) with the
linearization performed about the constant state,%
\begin{equation}
\left(  \overline{m},P,v,T\right)  =\left(  m_{\text{sl}},P_{\text{sl}%
},0,T_{\text{sl}}\right)  .\label{c129}%
\end{equation}
Taking the linearized gravitational force to be%
\begin{equation}
\overline{m}\left(  f_{\overline{M}}\right)  _{1}=-m_{\text{sl}}g,\label{c130}%
\end{equation}
one obtains%
\begin{align}
0  & =\frac{dv}{dx}\nonumber\\
0  & =\frac{d^{2}P}{dx^{2}}\label{c131}\\
0  & =-\frac{dP}{dx}+\eta_{\text{sl}}\frac{d^{2}v}{dx^{2}}-m_{\text{sl}%
}g\nonumber\\
0  & =\frac{d^{2}T}{dx^{2}}.\nonumber
\end{align}
We solve these equations with the boundary conditions,%
\begin{equation}
P\left(  0\right)  =P_{\text{sl}}\text{ and }T\left(  0\right)  =T_{\text{sl}%
},\label{c132}%
\end{equation}
to find%
\begin{align}
v\left(  x\right)   & =\mathcal{C}_{1}\nonumber\\
P\left(  x\right)   & =P_{\text{sl}}-m_{\text{sl}}gx\label{c133}\\
T\left(  x\right)   & =T_{\text{sl}}+\mathcal{C}_{2}x,\nonumber
\end{align}
where (\ref{c133} b) is the hydrostatic equation (\ref{c121}), and the
$\mathcal{C}$'s\ are constants we must find using extra constraints. \ The
$\overline{M}\left(  D,\eta\right)  $-formulation flux equations (\ref{s11})
and (\ref{s12}) for this one-dimensional problem become%
\begin{equation}
\left(  j_{E}\right)  _{1}=-D_{\text{sl}}\left\{
\begin{array}
[c]{c}%
\left[  m\left(  c_{P}-h_{M}\alpha_{P}\right)  \right]  _{\text{sl}}\frac
{dT}{dx}+\\
\left(  \frac{h_{M}\gamma}{c^{2}}-T\alpha_{P}\right)  _{\text{sl}}\frac
{dP}{dx}%
\end{array}
\right\}  +\left(  mh_{M}\right)  _{\text{sl}}v\label{c134}%
\end{equation}
and%
\begin{equation}
\left(  j_{M}\right)  _{1}=-D_{\text{sl}}\left[  -\left(  m\alpha_{P}\right)
_{\text{sl}}\frac{dT}{dx}+\left(  \frac{\gamma}{c^{2}}\right)  _{\text{sl}%
}\frac{dP}{dx}\right]  +m_{\text{sl}}v.\label{c135}%
\end{equation}
It makes physical sense to enforce the constraints that there are no mass or
total energy flux through the atmosphere due to gravity:%
\begin{equation}
\left(  j_{E}\right)  _{1}=\left(  j_{M}\right)  _{1}=0.\label{c136}%
\end{equation}
Therefore, using (\ref{c134}) and (\ref{c135}) together with solution
(\ref{c133}), these constraints imply%
\begin{align}
0  & =-D_{\text{sl}}\left\{
\begin{array}
[c]{c}%
\left[  m\left(  c_{P}-h_{M}\alpha_{P}\right)  \right]  _{\text{sl}%
}\mathcal{C}_{2}-\\
\left(  \frac{h_{M}\gamma}{c^{2}}-T\alpha_{P}\right)  _{\text{sl}}%
m_{\text{sl}}g
\end{array}
\right\}  +\left(  mh_{M}\right)  _{\text{sl}}\mathcal{C}_{1}\label{c137}\\
0  & =-D_{\text{sl}}\left[  -\left(  m\alpha_{P}\right)  _{\text{sl}%
}\mathcal{C}_{2}-\left(  \frac{\gamma}{c^{2}}\right)  _{\text{sl}}%
m_{\text{sl}}g\right]  +m_{\text{sl}}\mathcal{C}_{1}\nonumber
\end{align}
which, when solved simultaneously for the two constants, gives%
\begin{equation}
\mathcal{C}_{1}=-g\left(  \frac{D}{c^{2}}\right)  _{\text{sl}}\text{ and
}\mathcal{C}_{2}=-\left(  \frac{T\alpha_{P}}{c_{P}}\right)  _{\text{sl}%
}g,\label{c138}%
\end{equation}
where I have used thermodynamic relationship (\ref{e12.1}) to write
(\ref{c138} a). \ Thus, solution (\ref{c133}) becomes%
\begin{align}
v\left(  x\right)   & =-g\left(  \frac{D}{c^{2}}\right)  _{\text{sl}%
}\nonumber\\
P\left(  x\right)   & =P_{\text{sl}}-m_{\text{sl}}gx\label{c139}\\
T\left(  x\right)   & =T_{\text{sl}}-\left(  \frac{T\alpha_{P}}{c_{P}}\right)
_{\text{sl}}gx.\nonumber
\end{align}
Note the following: (1) thermodynamic relationship (\ref{e5.5}) implies%
\begin{equation}
\frac{d\sigma}{dx}=\left(  \frac{c_{P}}{T}\right)  _{\text{sl}}\frac{dT}%
{dx}-\left(  \frac{\alpha_{P}}{m}\right)  _{\text{sl}}\frac{dP}{dx}%
\label{c140}%
\end{equation}
and when (\ref{c139} b and c) are used in the right-hand side, \textit{this
yields the isentropic condition (\ref{c123})}, (2) when classical ideal gas
relationships (\ref{e26}) and (\ref{e27.5}) are used in (\ref{c139} c), one
finds the temperature lapse rate to be the same as the one predicted by
equation (\ref{c125}), and (3) using the estimates $c_{\text{sl}}=340$
m/s\ and $D_{\text{sl}}=1.82\times10^{-5}$ m$^{2}$/s\ for air at sea
level,\footnote{These values arise from assuming the average temperature and
pressure at sea level are $T_{\text{sl}}=288.15$ K and $P_{\text{sl}%
}=1.013\times10^{5}$ Pa. \ The sound speed $c_{\text{sl}}$ is computed from
the classical ideal gas formula (\ref{dc5})\ and the diffusion coefficient
$D_{\text{sl}}$ is found in \textsc{table} \ref{tab2}\ of \textsc{appendix
\ref{difp}}.} the above equation predicts a very small constant velocity
directed towards the Earth,%
\begin{equation}
v=-1.54\times10^{-9}\text{ m/s,}\label{c141}%
\end{equation}
which is analogous to the thermophoretic velocity treated in \S \ref{therm}.

Solving steady-state NSF equations (\ref{s13}) linearized about state
(\ref{c129}), with body force (\ref{c130}) and boundary conditions
(\ref{c132}), one obtains the same form of solution as (\ref{c133}). \ The
flux equations (\ref{s14}) and (\ref{s15}) for this one-dimensional problem
are%
\begin{equation}
\left(  j_{E}\right)  _{1}=-\left(  k_{F}\right)  _{\text{sl}}\frac{dT}%
{dx}+\left(  mh_{M}\right)  _{\text{sl}}v\label{c142}%
\end{equation}
and%
\begin{equation}
\left(  j_{M}\right)  _{1}=m_{\text{sl}}v,\label{c142.1}%
\end{equation}
and therefore we see that enforcing constraints (\ref{c136}) for no mass or
total energy flux, gives rise to a zero velocity solution and \textit{an
isothermal condition}.

\section{\label{poise}Poiseuille Flow}

\begin{figure}
[ptb]
\begin{center}
\includegraphics[
height=3.0493in,
width=4.5455in
]%
{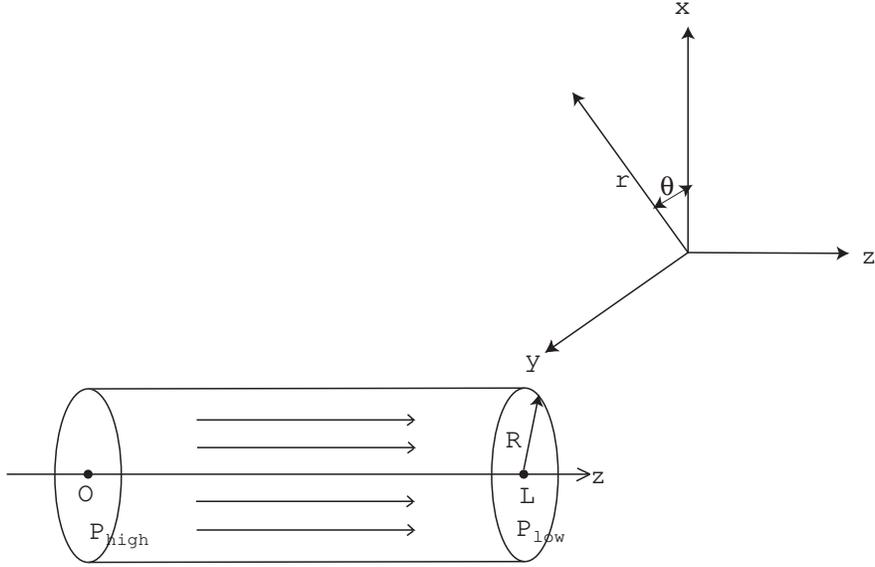}%
\caption{Poiseuille flow}%
\label{fig3}%
\end{center}
\end{figure}

So far in this paper, I have neglected to study any examples that result in a
non-zero rotational (divergence-free) part of the velocity, $\underline{v}%
_{R}$. \ Therefore, let us now look at a basic example in which this part of
the velocity arises: laminar Poiseuille flow through a cylinder, which is
depicted schematically in \textsc{figure \ref{fig3}}.%
Suppose that on the $z=0$\ side, the fluid is maintained at a pressure
$P_{\text{high}}$ and temperature $T_{\ast}$\ and on the $z=L$ side with a
lower pressure of $P_{\text{low}}$ and the same temperature $T_{\ast}$.
\ This, of course, causes steady-state flow in the $+z$-direction. \ In
addition, let us make the following assumptions: (1) the radius $\mathcal{R}$
of the cylinder\ is much larger than the mean free path length of the fluid so
that $K\!n$ is small, (2) the length $L$ of the cylinder\ is quite a bit
larger than the radius so that possible entrance and exit effects at the ends
may be neglected , (3) the pressure gradient is small so that we may
linearize, (4) parameters are chosen to yield a fairly small Reynolds number
so that the flow is laminar (i.e. the radius\ and the pressure gradient cannot
be too large and the shear viscosity cannot be too small), and (5) the
cylinder itself has uniform temperature $T_{\ast}$. \ It is clear that we can
model this as a two-dimensional cylindrical problem in the $r$\ and $z$\ variables.

Let $P_{\ast}$ represent the average pressure,%
\begin{equation}
P_{\ast}=\frac{P_{\text{high}}+P_{\text{low}}}{2}\label{c145}%
\end{equation}
and suppose everything with a subscript "$\ast$" is evaluated at $\left(
P_{\ast},T_{\ast}\right)  $. \ For this two-dimensional problem, the
steady-state $\overline{M}\left(  D,\eta\right)  $-formulation (\ref{s10})
with the linearization performed about%
\begin{equation}
\left(  \overline{m},P,\underline{v},T\right)  =\left(  m_{\ast},P_{\ast
},\underline{0},T_{\ast}\right)  ,\label{c144}%
\end{equation}
is expressed--with the use of the cylindrical coordinate formulas presented in
\textsc{appendix \ref{apptens}}--as%
\begin{align}
0  & =\frac{1}{r}\frac{\partial}{\partial r}\left(  rv_{r}\right)
+\frac{\partial v_{z}}{\partial z}\nonumber\\
0  & =\frac{1}{r}\frac{\partial}{\partial r}\left(  r\frac{\partial
P}{\partial r}\right)  +\frac{\partial^{2}P}{\partial z^{2}}\nonumber\\
0  & =-\frac{\partial P}{\partial r}+\eta_{\ast}\left\{  \frac{\partial
}{\partial r}\left[  \frac{1}{r}\frac{\partial}{\partial r}\left(
rv_{r}\right)  \right]  +\frac{\partial^{2}v_{r}}{\partial z^{2}}\right\}
\label{c147}\\
0  & =-\frac{\partial P}{\partial z}+\eta_{\ast}\left[  \frac{1}{r}%
\frac{\partial}{\partial r}\left(  r\frac{\partial v_{z}}{\partial r}\right)
+\frac{\partial^{2}v_{z}}{\partial z^{2}}\right] \nonumber\\
0  & =\frac{1}{r}\frac{\partial}{\partial r}\left(  r\frac{\partial
T}{\partial r}\right)  +\frac{\partial^{2}T}{\partial z^{2}},\nonumber
\end{align}
which we intend to solve with the following boundary conditions:%
\begin{align}
\left.  P\right\vert _{z=0}  & =P_{\text{high}}\label{c149}\\
\left.  P\right\vert _{z=L}  & =P_{\text{low}},\label{c149.1}%
\end{align}%
\begin{equation}
\left.  T\right\vert _{z=0}=\left.  T\right\vert _{z=L}=\left.  T\right\vert
_{r=\mathcal{R}}=T_{\ast},\label{c150}%
\end{equation}
the no-slip condition,\footnote{These are appropriate under our small $K\!n$
conditions.}%
\begin{equation}
\left.  v_{z}\right\vert _{r=\mathcal{R}}=0,\label{c151}%
\end{equation}
and the requirement that%
\begin{equation}
\left.  v_{z}\right\vert _{r=0}\text{ is finite.}\label{c151.1}%
\end{equation}
Immediately, we see that solving Laplace's equation (\ref{c147} e) with
boundary conditions (\ref{c150}) for the temperature, yields the constant
solution,%
\begin{equation}
T=T_{\ast}.\label{c152.1}%
\end{equation}
With this fact and cylindrical coordinate formulas (\ref{t1}) and (\ref{t3}),
one may express the $r$ and $z$ components of the $\overline{M}\left(
D,\eta\right)  $-formulation total energy and mass fluxes (\ref{s11}) and
(\ref{s12}) as%
\begin{align}
\left(  j_{E}\right)  _{r}  & =-\left[  D\left(  \frac{h_{M}\gamma}{c^{2}%
}-T\alpha_{P}\right)  \right]  _{\ast}\frac{\partial P}{\partial r}+\left(
mh_{M}\right)  _{\ast}v_{r}\label{c152.2}\\
\left(  j_{E}\right)  _{z}  & =-\left[  D\left(  \frac{h_{M}\gamma}{c^{2}%
}-T\alpha_{P}\right)  \right]  _{\ast}\frac{\partial P}{\partial z}+\left(
mh_{M}\right)  _{\ast}v_{z}\label{c152.3}%
\end{align}
and%
\begin{align}
\left(  j_{M}\right)  _{r}  & =-\left[  D\left(  \frac{\gamma}{c^{2}}\right)
\right]  _{\ast}\frac{\partial P}{\partial r}+m_{\ast}v_{r}\label{c152.4}\\
\left(  j_{M}\right)  _{z}  & =-\left[  D\left(  \frac{\gamma}{c^{2}}\right)
\right]  _{\ast}\frac{\partial P}{\partial z}+m_{\ast}v_{z}.\label{c152.5}%
\end{align}
Next, one observes that for this steady-state problem to be physical, there
must be no radial mass or total energy flux, i.e.%
\begin{equation}
\left(  j_{E}\right)  _{r}=\left(  j_{M}\right)  _{r}=0\label{c152.6}%
\end{equation}
must be satisfied, which requires%
\begin{equation}
\frac{\partial P}{\partial r}=0\text{ and }v_{r}=0.\label{c152.7}%
\end{equation}
Substituting (\ref{c152.7}) into equations (\ref{c147} a-d) reduces them to%
\begin{align}
0  & =\frac{\partial v_{z}}{\partial z}\nonumber\\
0  & =\frac{\partial^{2}P}{\partial z^{2}}\label{c152.8}\\
0  & =-\frac{\partial P}{\partial z}+\eta_{\ast}\left[  \frac{1}{r}%
\frac{\partial}{\partial r}\left(  r\frac{\partial v_{z}}{\partial r}\right)
+\frac{\partial^{2}v_{z}}{\partial z^{2}}\right]  ,\nonumber
\end{align}
and since (\ref{c152.8} a) implies that $v_{z}$\ is a function of $r$ only, we
are left to solve%
\begin{align}
0  & =\frac{\partial^{2}P}{\partial z^{2}}\label{c152.9}\\
0  & =-\frac{\partial P}{\partial z}+\eta_{\ast}\left[  \frac{1}{r}%
\frac{\partial}{\partial r}\left(  r\frac{\partial v_{z}}{\partial r}\right)
\right] \label{c152.91}%
\end{align}
with boundary conditions (\ref{c149}), (\ref{c149.1}), (\ref{c151}), and
(\ref{c151.1}). \ Doing so, yields the familiar solution,%
\begin{align}
P\left(  z\right)   & =P_{\text{high}}-\left(  \frac{P_{\text{high}%
}-P_{\text{low}}}{L}\right)  z\label{c152.92}\\
v_{z}\left(  r\right)   & =\left(  \frac{P_{\text{high}}-P_{\text{low}}}%
{4\eta_{\ast}L}\right)  \left(  \mathcal{R}^{2}-r^{2}\right)  ,\label{c152.93}%
\end{align}
which is the same as the one predicted by the NSF equations--see Bird,
Stewart, and Lightfoot \cite[pp. 48-52]{bsl}. \ However, it is important to
remark that in the NSF description, the linear equation (\ref{c152.92}) for
the pressure is \textit{assumed} rather than derived.\footnote{This is
because--unlike the $\overline{M}\left(  D,\eta\right)  $-formulation--the NSF
steady-state description lacks Laplace's equation for the pressure (\ref{s10}
b).}

Next, the average velocity is computed as%
\begin{equation}
\overline{v}_{z}=\frac{\int\nolimits_{0}^{2\pi}\int\nolimits_{0}^{\mathcal{R}%
}v_{z}\left(  r\right)  rdrd\theta}{\int\nolimits_{0}^{2\pi}\int
\nolimits_{0}^{\mathcal{R}}rdrd\theta}\label{c153}%
\end{equation}
and the average convective mass flow rate as%
\begin{equation}
\dot{M}_{\text{conv}}=\pi\mathcal{R}^{2}m_{\text{av}}\overline{v}%
_{z},\label{c154}%
\end{equation}
which evaluates to the familiar Hagen-Poiseuille formula predicted by the NSF
equations,\footnote{See Bird, Stewart, and Lightfoot \cite[p. 51]{bsl}.}%
\begin{equation}
\dot{M}_{\text{conv}}=\frac{\pi\mathcal{R}^{4}}{8}\left(  \frac{m}{\eta
}\right)  _{\ast}\left(  \frac{P_{\text{high}}-P_{\text{low}}}{L}\right)
,\label{c155}%
\end{equation}
when velocity profile (\ref{c152.93}) is used. \ However, in the $\overline
{M}\left(  D,\eta\right)  $-formulation, there is an additional diffusive
contribution to the mass flow rate, given by%
\begin{equation}
\dot{M}_{\text{dif}}=\pi\mathcal{R}^{2}\left(  q_{M}\right)  _{z},\label{c156}%
\end{equation}
where the non-convective mass flux $\left(  q_{M}\right)  _{z}$\ is found by
subtracting the convective part from (\ref{c152.5}) and using pressure
equation (\ref{c152.92}), which yields%
\begin{equation}
\dot{M}_{\text{dif}}=\pi\mathcal{R}^{2}\left(  \frac{D\gamma}{c^{2}}\right)
_{\ast}\left(  \frac{P_{\text{high}}-P_{\text{low}}}{L}\right)  .\label{c166}%
\end{equation}
The total mass flow rate for the $\overline{M}\left(  D,\eta\right)
$-formulation is then the sum of the convective and diffusive parts:%
\begin{equation}
\dot{M}=\left(  \frac{P_{\text{high}}-P_{\text{low}}}{L}\right)
\pi\mathcal{R}^{2}\left[  \frac{\mathcal{R}^{2}}{8}\left(  \frac{m}{\eta
}\right)  _{\ast}+\left(  \frac{D\gamma}{c^{2}}\right)  _{\ast}\right]
.\label{c167}%
\end{equation}
Note that under the assumptions given at the beginning of this section the
diffusive part is very small compared to the convective part, and so the flow
is enhanced only slightly by this term in the hydrodynamic regime, e.g. for
$\mathcal{R}=1$ mm and air close to normal temperature and pressure, $\dot
{M}_{\text{conv}}$ is found to be on the order of $10^{7}$ times larger than
$\dot{M}_{\text{dif}}$. \ Let us more generally remark that in all problems
for which\ the rotational part of the velocity dominates, one finds that
$\overline{M}\left(  D,\eta\right)  $-formulation and NSF predictions barely
differ from one another in the hydrodynamic regime--and sometimes the
predictions are identical, as in the case of Couette flow. \ Since viscometers
are based on these types of phenomena, this provides justification for our
assumption that the shear viscosity appearing in the $\overline{M}\left(
D,\eta\right)  $-formulation may, for all intents and purposes, be taken to be
the same as the Navier-Stokes shear viscosity.

As one final observation, when ideal gas formulas (\ref{e29}) and (\ref{e30})
are used, the diffusive mass flow rate (\ref{c166}) becomes%
\begin{equation}
\dot{M}_{\text{dif}}=\pi\mathcal{R}^{2}\frac{A}{R}\left(  \frac{D}{T}\right)
_{\ast}\left(  \frac{P_{\text{high}}-P_{\text{low}}}{L}\right)  ,\label{c168}%
\end{equation}
which has the same form as the one that Veltzke and Th\"{o}ming introduce in
\cite[p. 413 eq. (3.19)]{veltz} via Fick's law.

\section{\label{therm}Thermophoresis}

\begin{figure}
[ptb]
\begin{center}
\includegraphics[
height=4.3708in,
width=4.5455in
]%
{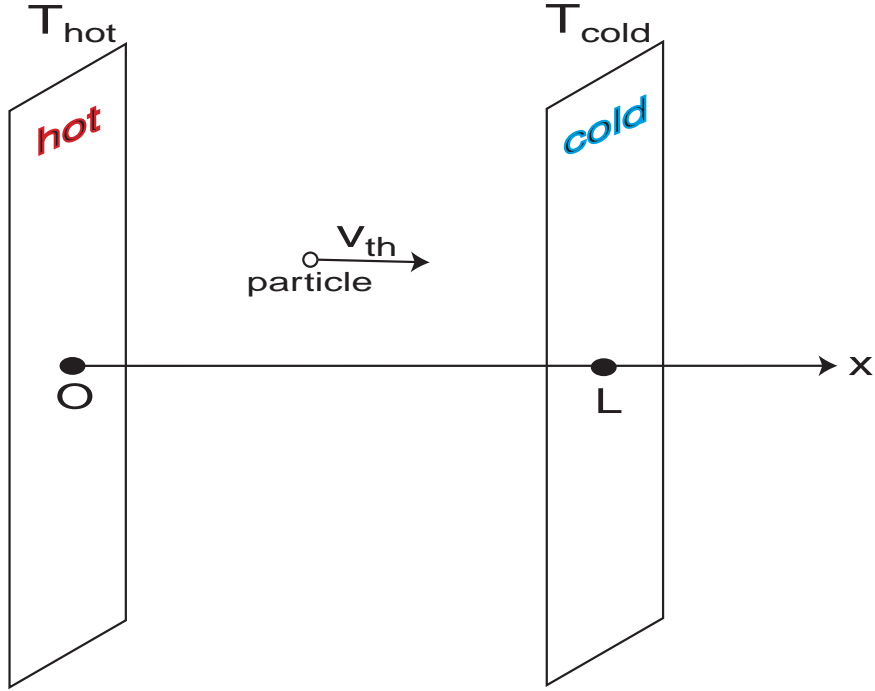}%
\caption{thermophoresis}%
\label{fig4}%
\end{center}
\end{figure}

When a macroscopic particle is placed in a resting fluid with a temperature
gradient, the particle tends to move in the cooler direction. \ This
phenomenon is known as thermophoresis, and as Brenner \cite{brenth2} points
out, it is modelled with the NSF formulation only by invoking a
molecularly-based thermal slip boundary condition. \ However, the
$\overline{M}\left(  D,\eta\right)  $-formulation--and others, like Brenner's,
which feature a non-convective mass flux--can provide a natural mechanism for
thermophoresis even when a no-slip boundary condition is applied at the
particle surface. \ The simplest problem that may be associated with
thermophoresis is examined below using the $\overline{M}\left(  D,\eta\right)
$-formulation. \ In it, gravity and possible boundary layers are neglected,
and only small temperature gradients are considered. \ Also, the
thermophoretic particle is assumed to be insulated and to have a radius, $R$,
much larger than the fluid's mean free path length so that the Knudsen number
is small. \ As shown schematically in \textsc{figure} \ref{fig4}, we suppose
that a fluid with average pressure $P_{\ast}$ lies between two parallel
plates, a hotter plate at $x=0$ maintained at temperature, $T_{\text{hot}}%
$,\ and a colder plate at $x=L$ kept at temperature, $T_{\text{cold}}$.
\ These plates are assumed to be impermeable.%

This is another steady-state Cartesian one-dimensional problem with variation
in the $x_{1}\equiv x$ direction only and $v_{1}\equiv v$ as the only non-zero
velocity component. \ Therefore, the $\overline{M}\left(  D,\eta\right)
$-formulation (\ref{s10}) linearized about the constant state,%
\begin{equation}
\left(  \overline{m},P,v,T\right)  =\left(  m_{\ast},P_{\ast},0,T_{\ast
}\right)  ,\label{c169}%
\end{equation}
where $T_{\ast}$ is the average temperature,%
\begin{equation}
T_{\ast}=\frac{T_{\text{hot}}+T_{\text{cold}}}{2},\label{c170}%
\end{equation}
becomes%
\begin{align}
0  & =\frac{dv}{dx}\nonumber\\
0  & =\frac{d^{2}P}{dx^{2}}\label{c171}\\
0  & =-\frac{dP}{dx}+\eta_{\ast}\frac{d^{2}v}{dx^{2}}\nonumber\\
0  & =\frac{d^{2}T}{dx^{2}}.\nonumber
\end{align}
Assuming the temperature of the fluid at the plates equals that of the plates,
gives rise to the boundary conditions,%
\begin{equation}
\left.  T\right\vert _{x=0}=T_{\text{hot}}\text{ and }\left.  T\right\vert
_{x=L}=T_{\text{cold}},\label{c172}%
\end{equation}
and using these with equation (\ref{c171} d) yields the linear temperature
profile,%
\begin{equation}
T\left(  x\right)  =T_{\text{hot}}-\left(  \frac{T_{\text{hot}}-T_{\text{cold}%
}}{L}\right)  x.\label{c173}%
\end{equation}
Furthermore, equations (\ref{c171} a and c) imply that $v$\ and $P$\ are both
constants. \ Therefore,%
\begin{equation}
P=P_{\ast},\label{c174}%
\end{equation}
and since the plates are impermeable, we may find $v$\ by enforcing the no
mass flux condition,\footnote{Note that in this case, it is not appropriate to
enforce a no total energy flux condition, since there is heat being put into
one side and taken out of the other.}%
\begin{equation}
\left(  j_{M}\right)  _{1}=0,\label{c175}%
\end{equation}
where employing (\ref{c173}) and (\ref{c174}) in the one-dimensional version
of (\ref{s12}), we have%
\begin{equation}
\left(  j_{M}\right)  _{1}=-\left(  m\alpha_{P}D\right)  _{\ast}\left(
\frac{T_{\text{hot}}-T_{\text{cold}}}{L}\right)  +m_{\ast}v.\label{c176}%
\end{equation}
The constant velocity is then%
\begin{equation}
v=\left(  \alpha_{P}D\right)  _{\ast}\left(  \frac{T_{\text{hot}%
}-T_{\text{cold}}}{L}\right)  .\label{c177}%
\end{equation}

Therefore, my theory predicts a constant velocity in the $+x$ (colder)
direction that is proportional to the fluid's diffusion coefficient and
thermal expansion coefficient and the size of the temperature gradient imposed
between the plates. \ If an insulated macroscopic particle were present in
such a velocity field with a no-slip condition applied at its surface and no
other forces acting on it, then the particle would simply be carried along at
the fluid's velocity, i.e. the particle's thermophoretic velocity
$v_{\text{th}}$ would equal the fluid's continuum velocity $v$ computed
above.\footnote{This is, of course, a gross oversimplification of the real
problem of thermophoresis in that we have neglected any and all properties of
the particle.}

In contrast, when the non-convective mass flux is assumed to be zero, as in
the case of the NSF formulation, the no mass flow condition (\ref{c175})
implies%
\begin{equation}
v=0.\label{c115}%
\end{equation}
Consequently, thermophoretic motion does not arise in the same sense that it
does for the $\overline{M}\left(  D,\eta\right)  $-formulation. \ To account
for thermophoresis with the NSF equations, one then finds it necessary to
employ a thermal slip boundary condition at the particle's surface at which
there is assumed to be a non-zero tangential velocity proportional to the
temperature gradient and in the down-gradient direction. \ Under this
assumption, various researchers have been led, theoretically and
experimentally, to the following expression for the thermophoretic velocity in
the case of an insulated particle in a gas:%
\begin{equation}
\underline{v}_{\text{th}}=C_{\text{th}}\frac{\eta}{mT}\nabla T,\label{c116}%
\end{equation}
where $C_{\text{th}}$\ is the thermal slip coefficient, dimensionless and
theoretically believed to lie in the interval, $0.75\leq C_{\text{th}}\leq
1.5$, with $C_{\text{th}}\approx1.17$ being a widely accepted theoretical
estimate for gases.\footnote{See Brenner and Bielenberg \cite{brenth1},
Brenner \cite{brenth2}, and Derjaguin et al. \cite{der}.} \ For the problem
considered here, (\ref{c116}) becomes%
\begin{equation}
v_{\text{th}}=C_{\text{th}}\left(  \frac{\eta}{mT}\right)  _{\ast}\left(
\frac{T_{\text{hot}}-T_{\text{cold}}}{L}\right)  .\label{c117}%
\end{equation}

Employing (I.72), which implies%
\begin{equation}
\left(  mD\right)  _{\ast}=\left(  C\eta\right)  _{\ast},\label{c117.3}%
\end{equation}
and assuming an ideal gas for which the thermal expansion coefficient is given
by (\ref{e26}), the thermophoretic velocity predicted by my theory in equation
(\ref{c177}) may be expressed as%
\begin{equation}
v_{\text{th}}=\left(  C\frac{\eta}{mT}\right)  _{\ast}\left(  \frac
{T_{\text{hot}}-T_{\text{cold}}}{L}\right)  .\label{c118}%
\end{equation}
Therefore, by identifying $C_{\ast}$\ with $C_{\text{th}}$, one finds this to
be \textit{the same as the thermal slip formula} (\ref{c117}). \ Also, recall
that in \S \ref{sound}, the value $C=7/6\approx1.17$\ was obtained for a
classical monatomic ideal gas.\footnote{Values of $C$ pertaining to a
selection of other gases are provided in \textsc{appendix} \ref{difp}.}

As discussed in Brenner \cite{brenth2}, there is no real theory for thermal
slip in a liquid. \ Therefore, researchers, e.g. McNab and Meisen
\cite{mcnab}, typically apply the gas formulas to liquids, even though they
admit that there is no theoretical justification for doing so. \ For the
idealized problem considered here, McNab and Meisen's data corresponds to a
thermal slip coefficient for water and $n$-hexane near room temperature of
$C_{\text{th}}\approx0.13$, an order of magnitude smaller than that of a gas.
\ Again using (\ref{c117.3}), but no longer assuming the ideal gas
relationship (\ref{e26}), my theory's thermophoretic velocity (\ref{c177})
becomes%
\begin{equation}
v_{\text{th}}=-\left(  C\frac{\eta\alpha_{P}}{m}\right)  _{\ast}\left(
\frac{T_{\text{hot}}-T_{\text{cold}}}{L}\right)  ,\label{c119}%
\end{equation}
which when compared with the thermal slip formula (\ref{c117}), yields the
relationship,%
\begin{equation}
C_{\text{th}}=\left(  CT\alpha_{P}\right)  _{\ast}.\label{c120}%
\end{equation}
For water at $T_{\ast}=293$ K, we take $C_{\ast}=2.1$ (see \textsc{appendix
\ref{difp}}) and $\left(  \alpha_{P}\right)  _{\ast}=2.05\times10^{-4}$
K$^{-1}$ (from the CRC Handbook of Chemistry and Physics \cite{CRC}), giving a
slip coefficient value of $C_{\text{th}}=0.13$, which is \textit{the same as
the experimental value mentioned above}.

It is interesting to note that there is, in general, no positive definite
restriction on the thermal expansion coefficient $\alpha_{P}$. \ There exist
equilibrium thermodynamic states in water and liquid helium, for example, in
which $\alpha_{P}$\ may vanish or become negative. \ In view of equation
(\ref{c120}), this means that such states could give rise to the occurrence of
no thermophoresis or reverse thermophoresis.

\section{\label{shock}Shock Waves}

The problem examined below strains the limits of what one should expect from
the NSF and $\overline{M}\left(  D,\eta\right)  $ formulations in two ways:
(1) it is not near-equilibrium, which is counter to the assumption made in
\S 4\ and \textsc{appendix B}\ of \textsc{part I}\ used to justify linear
constitutive laws and (2) it is a $K\!n\sim O\left(  1\right)  $\ problem and,
therefore, not in the hydrodynamic regime. \ Nonetheless, the NSF equations
are commonly employed as a qualitative tool for investigating the internal
structure of shock waves, and so it is of interest to see how predictions from
the $\overline{M}\left(  D,\eta\right)  $-formulation compare with these.

\begin{figure}
[ptb]
\begin{center}
\includegraphics[
height=3.301in,
width=4.5455in
]%
{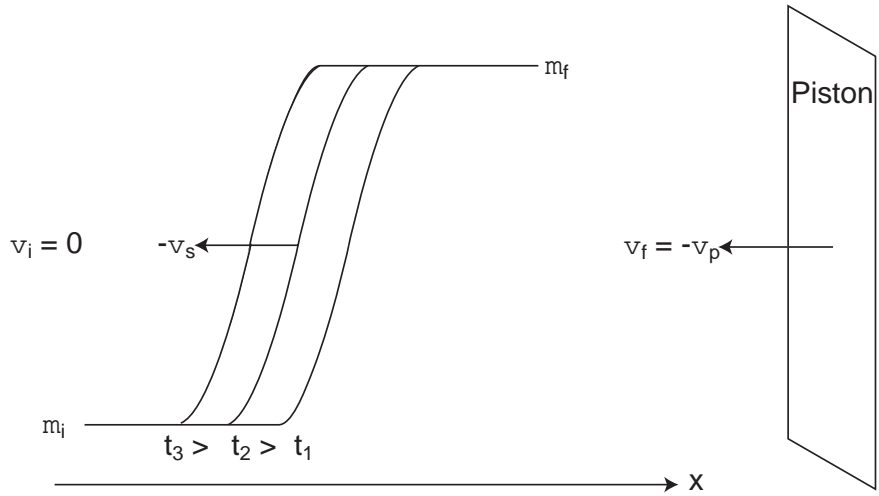}%
\caption{shock wave in a fixed coordinate system}%
\label{fig5}%
\end{center}
\end{figure}

\begin{figure}
[ptb]
\begin{center}
\includegraphics[
height=2.2857in,
width=3.5405in
]%
{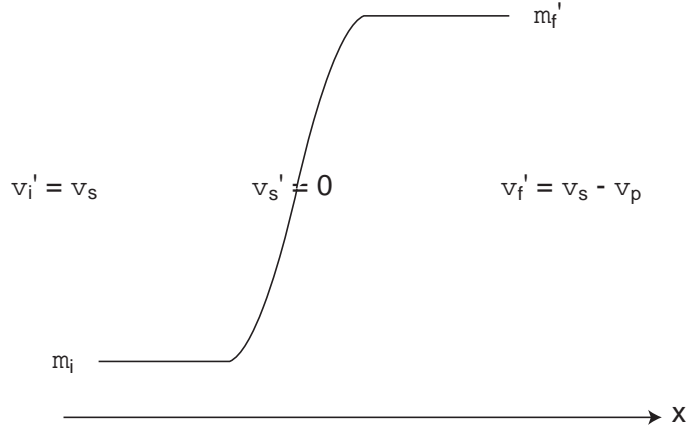}%
\caption{shock wave in a coordinate system moving with the front}%
\label{fig6}%
\end{center}
\end{figure}

Let us consider a Cartesian one-dimensional steady-state shock wave like the
one described in Zel'dovich and Raizer \cite[pp. 469-482]{zel} and depicted
schematically in \textsc{figure} \ref{fig5} in a fixed reference frame.
\ Again, we assume variation is in the $x_{1}\equiv x$ direction only,
$v_{1}\equiv v$\ is the only non-zero component of the velocity, and there are
no body forces. \ The subscript "$i$" is used to indicate undisturbed initial
values ahead of the shock, and the subscript "$f$" is used for final values
after the shock moves through. \ The initial and final fluid velocities are
taken to be%
\begin{equation}
v_{i}=0\text{ and }v_{f}=-v_{p},\label{c76}%
\end{equation}
where $v_{p}$\ is the constant speed of the piston generating the shock wave.
\ For convenience, one transforms the problem into a coordinate system moving
with the shock wave front as illustrated in \textsc{figure} \ref{fig6}, where
primes denote variables in the moving coordinate system and $v_{s}$ represents
the constant speed of the shock front.%

Expressing the non-linear balance laws (\ref{s1} a-c) and (\ref{s5})--with
constitutive equations (\ref{s2}), no body forces, and (I.18), (I.13), and
$w=0$--in the one-dimensional moving coordinate system, making the
steady-state assumption that the time derivatives vanish, using (\ref{e.6})
and (\ref{e.7}) to cast the system in terms of the variables, $\overline
{m}^{\prime}$, $m^{\prime}$, $v^{\prime}$, and $T^{\prime}$, and employing the
equilibrium thermodynamic relationships (\ref{e25})-(\ref{e27}), (\ref{e32}),
and (\ref{e33}) for a classical monatomic ideal gas, yields the following
system of ordinary differential equations:%
\begin{align}
\frac{d}{dx^{\prime}}\left(  \overline{m}^{\prime}v^{\prime}\right)   &
=0\label{c78}\\
\frac{d}{dx^{\prime}}\left(  -D^{\prime}\frac{dm^{\prime}}{dx^{\prime}%
}+m^{\prime}v^{\prime}\right)   & =0\label{c79}\\
\frac{d}{dx^{\prime}}\left(  \frac{RT^{\prime}m^{\prime}}{A}-\overline
{m}^{\prime}D^{\prime}\frac{dv_{x}^{\prime}}{dx^{\prime}}+\overline{m}%
^{\prime}v^{\prime2}\right)   & =0\label{c80}\\
\frac{d}{dx^{\prime}}\left[
\begin{array}
[c]{c}%
-\frac{3R}{2A}D^{\prime}\frac{d\left(  m^{\prime}T^{\prime}\right)
}{dx^{\prime}}-\overline{m}^{\prime}D^{\prime}\frac{dv^{\prime}}{dx^{\prime}%
}v^{\prime}+\\
\left(  \frac{5RT^{\prime}m^{\prime}}{2A\overline{m}^{\prime}}+\frac{1}%
{2}v^{\prime2}\right)  \overline{m}^{\prime}v^{\prime}%
\end{array}
\right]   & =0.\label{c81}%
\end{align}

For boundary conditions, let us assume that%
\begin{equation}
\lim_{x^{\prime}\rightarrow-\infty}\overline{m}^{\prime}=m_{i}\text{, }%
\lim_{x^{\prime}\rightarrow-\infty}m^{\prime}=m_{i}\text{, }\lim_{x^{\prime
}\rightarrow-\infty}v^{\prime}=v_{s}\text{, }\lim_{x^{\prime}\rightarrow
-\infty}T^{\prime}=T_{i}\label{c82}%
\end{equation}
and%
\begin{equation}
\lim_{x^{\prime}\rightarrow\infty}\overline{m}^{\prime}=\overline{m}%
_{f}\text{, }\lim_{x^{\prime}\rightarrow\infty}m^{\prime}=m_{f}\text{, }%
\lim_{x^{\prime}\rightarrow\infty}v^{\prime}=v_{s}-v_{p}\text{, }%
\lim_{x^{\prime}\rightarrow\infty}T^{\prime}=T_{f}\label{c83}%
\end{equation}
and that all gradients, $d\phi^{\prime}/dx^{\prime}$, vanish ahead of and
behind the shock wave front, i.e.%
\begin{equation}
\lim_{x^{\prime}\rightarrow-\infty}\frac{d\phi^{\prime}}{dx^{\prime}}%
=\lim_{x^{\prime}\rightarrow\infty}\frac{d\phi^{\prime}}{dx^{\prime}%
}=0,\label{c84}%
\end{equation}
where $\phi^{\prime}$\ may represent any of the variables, $\overline
{m}^{\prime}$, $m^{\prime}$, $v^{\prime}$, or $T^{\prime}$. \ Then, by
integrating (\ref{c78})-(\ref{c81}) from $-\infty$\ to $x^{\prime}$, one finds%
\begin{align}
\overline{m}^{\prime}v^{\prime}  & =m_{i}v_{s}\label{c85}\\
D^{\prime}\frac{dm^{\prime}}{dx^{\prime}}  & =\left(  \frac{m^{\prime}%
}{\overline{m}^{\prime}}-1\right)  m_{i}v_{s}\label{c86}\\
\overline{m}^{\prime}D^{\prime}\frac{dv^{\prime}}{dx^{\prime}}  & =\frac{R}%
{A}\left(  T^{\prime}m^{\prime}-T_{i}m_{i}\right)  +\left(  v^{\prime}%
-v_{s}\right)  m_{i}v_{s}\label{c87}\\
\frac{3R}{2A}D^{\prime}\frac{d\left(  m^{\prime}T^{\prime}\right)
}{dx^{\prime}}  & =\left[
\begin{array}
[c]{c}%
-D^{\prime}\frac{dv^{\prime}}{dx^{\prime}}+\\
\left(  \frac{5RT^{\prime}m^{\prime}}{2A\overline{m}^{\prime}}-\frac{5RT_{i}%
}{2A}\right)  +\\
\frac{1}{2}\left(  v^{\prime2}-v_{s}^{2}\right)
\end{array}
\right]  m_{i}v_{s},\label{c88}%
\end{align}
where (\ref{c85}) has been employed in writing equations (\ref{c86}%
)-(\ref{c88}), and by taking the limit of the above equations as $x^{\prime
}\rightarrow\infty$, one arrives at the relations%
\begin{gather}
\overline{m}_{f}=\frac{m_{i}v_{s}}{\left(  v_{s}-v_{p}\right)  }\label{c89}\\
m_{f}=\overline{m}_{f}\label{c90}\\
T_{f}=\frac{\left(  v_{s}-v_{p}\right)  }{v_{s}}\left(  T_{i}+\frac{A}{R}%
v_{s}v_{p}\right) \label{c91}\\
v_{s}^{2}-\frac{4}{3}v_{s}v_{p}-\frac{5RT_{i}}{3A}=0,\label{c92}%
\end{gather}
where (\ref{c89}) and (\ref{c90}) have been used to write (\ref{c91}), and
(\ref{c90}) and (\ref{c91}) have been used to obtain (\ref{c92}). \ One solves
(\ref{c92}) for the positive root to find the following expression for the
shock speed:%
\begin{equation}
v_{s}=\frac{2}{3}\left(  v_{p}+\sqrt{v_{p}^{2}+\frac{15RT_{i}}{4A}}\right)
.\label{c93}%
\end{equation}
Therefore, if the initial mass density and temperature, $m_{i}$\ and $T_{i}$,
and the piston speed, $v_{p}$,\ are known, then all of the relevant final
values, $\overline{m}_{f}$, $m_{f}$, $T_{f}$, and $v_{f}^{\prime}=v_{s}-v_{p}%
$, may be computed via (\ref{c93}) and (\ref{c89})-(\ref{c91}) for a classical
monatomic ideal gas. \ Note that these are identical to the shock speed and
final values found with the NSF equations.

Using the same assumptions and carrying out the foregoing procedure for the
NSF formulation, yields the following system:%
\begin{align}
m^{\prime}v^{\prime}  & =m_{i}v_{s}\label{c94}\\
\frac{4}{3}\eta^{\prime}\frac{dv^{\prime}}{dx^{\prime}}  & =\frac{R}{A}\left(
T^{\prime}m^{\prime}-T_{i}m_{i}\right)  +\left(  v^{\prime}-v_{s}\right)
m_{i}v_{s}\label{c95}\\
\frac{15R}{4A}\eta^{\prime}\frac{dT^{\prime}}{dx^{\prime}}  & =\left[
\begin{array}
[c]{c}%
-\frac{4\eta^{\prime}}{3m^{\prime}}\frac{dv^{\prime}}{dx^{\prime}}+\\
\left(  \frac{5RT^{\prime}}{2A}-\frac{5RT_{i}}{2A}\right)  +\\
\frac{1}{2}\left(  v^{\prime2}-v_{s}^{2}\right)
\end{array}
\right]  m_{i}v_{s}\label{c96}%
\end{align}
with boundary conditions (\ref{c82} b-d), (\ref{c83} b-d), and (\ref{c84}).
\ In the above, the classical monatomic ideal gas assumptions (\ref{c34}) and
(\ref{e29}) have been used.

For this non-linear problem, the longitudinal diffusion coefficient,
$D^{\prime}$,\ and the shear viscosity, $\eta^{\prime}$, may depend on state
variables, e.g. it is common to express the viscosity with a temperature power
law of the form (\ref{dc1}):
\[
\eta\left(  T\right)  =\eta_{r}\left(  \frac{T}{T_{r}}\right)  ^{\beta},
\]
where $T_{r}$\ is some reference temperature and $\eta_{r}$\ is the viscosity
measured at $T_{r}$ (see \textsc{appendix \ref{difp}}). \ Also, in
\textsc{appendix \ref{difp}}, it is shown that in a classical monatomic ideal
gas one might reasonably take (\ref{dc4}):%
\[
\left(  \overline{m}D\right)  \left(  T\right)  =\frac{7}{6}\eta_{r}\left(
\frac{T}{T_{r}}\right)  ^{\beta}%
\]

\begin{figure}
[ptb]
\begin{center}
\includegraphics[
height=3.0415in,
width=4.0421in
]%
{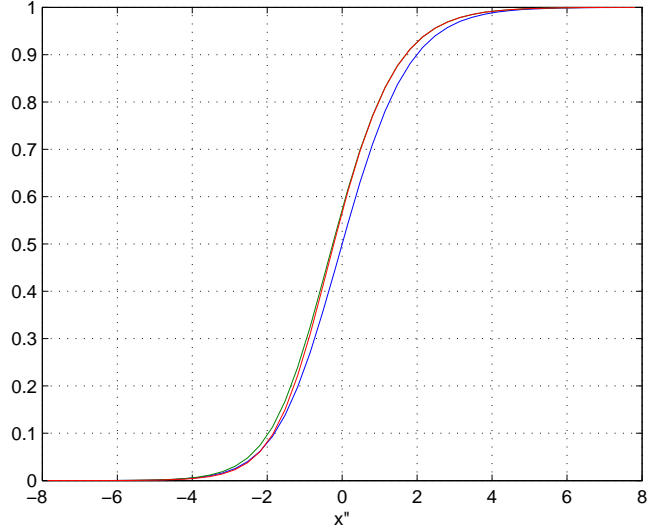}%
\caption{numerically computed shock wave profiles: $\overline{M}\left(
D,\eta\right)  $ - formulation}%
\label{fig7}%
\end{center}
\end{figure}

\begin{figure}
[ptb]
\begin{center}
\includegraphics[
height=2.7614in,
width=3.3399in
]%
{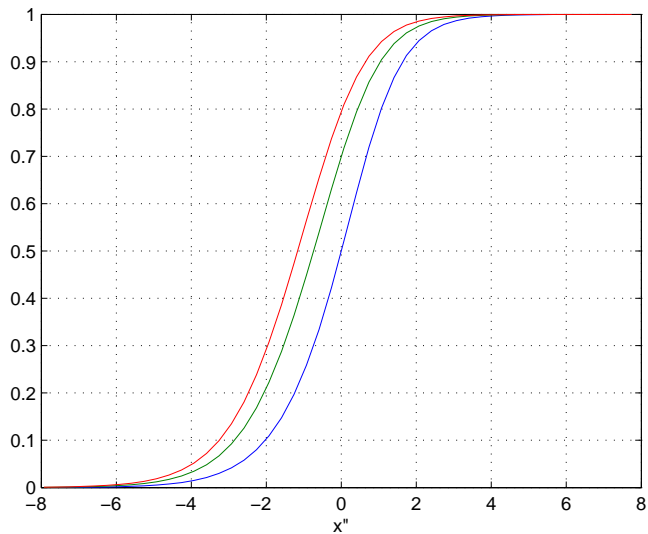}%
\caption{numerically computed shock wave profiles: NSF formulation}%
\label{fig8}%
\end{center}
\end{figure}

\textsc{Figures} \ref{fig7} and \ref{fig8}\ contain plots of numerical
solutions to dimensionless versions of (\ref{c85})-(\ref{c88}) and
(\ref{c94})-(\ref{c96}) with their appropriate boundary conditions and power
laws (\ref{dc1}) and (\ref{dc4}) describing the transport coefficients. \ The
numerical method employed is iterative Newton's method with centered
differences approximating the spatial derivatives. \ The dimensionless
variables used are%
\begin{equation}
x^{\prime\prime}=\frac{x^{\prime}}{\lambda_{i}}\text{, }v^{\prime\prime}%
=\frac{v^{\prime}}{c_{i}}\text{, }\overline{m}^{\prime\prime}=\frac
{\overline{m}^{\prime}}{m_{i}}\text{, }m^{\prime\prime}=\frac{m^{\prime}%
}{m_{i}}\text{, and }T^{\prime\prime}=\frac{T^{\prime}}{T_{i}},\label{c99}%
\end{equation}
where $\lambda_{i}$\ and $c_{i}$\ are the mean free path length and adiabatic
sound speed in the initial state, and the Mach number is computed as%
\begin{equation}
M\!a=\frac{v_{s}}{c_{i}}.\label{c100}%
\end{equation}
Each of the figures corresponds to $M\!a=2.05$ with parameters chosen to be
those of Alsmeyer \cite{als} for his shock wave experiments in argon gas: the
undisturbed initial values,%
\begin{align*}
T_{i}  & =300\text{ K}\\
m_{i}  & =1.07\times10^{-4}\text{ kg/m}^{3}\text{ (for }P_{i}=6.67\text{
Pa)}\\
\lambda_{i}  & =1.10\times10^{-3}\text{ m}\\
c_{i}  & =323\text{ m/s}%
\end{align*}
and the parameters appearing in (\ref{dc1}),%
\[
T_{r}=300\text{ K, }\eta_{r}=2.31\times10^{-5}\text{ kg/(m}\cdot\text{s), and
}\beta=0.72.
\]
The curves plotted in \textsc{figures} \ref{fig7} and \ref{fig8} are the
normalized mass density, $m^{\prime\prime\prime}=\left(  m^{\prime}%
-m_{i}\right)  /\left(  m_{f}-m_{i}\right)  $ (in blue), the normalized fluid
speed, $v^{\prime\prime\prime}=\left(  v^{\prime}-v_{i}^{\prime}\right)
/(v_{f}^{\prime}-v_{i}^{\prime})$ (in green), and the normalized temperature,
$T^{\prime\prime\prime}=\left(  T^{\prime}-T_{i}\right)  /\left(  T_{f}%
-T_{i}\right)  $ (in red) versus the dimensionless position, $x^{\prime\prime
}$. \ Since the numerical solution sets are arbitrarily positioned on the
$x^{\prime\prime}$-axis, the $x^{\prime\prime}=0$ point is chosen to be the
center of the mass density profile (the point at which the highest slope of
$m^{\prime\prime}$ occurs). \ \textsc{Figure} \ref{fig7} contains the
normalized shock wave profiles predicted by the $\overline{M}\left(
D,\eta\right)  $-formulation and \textsc{figure} \ref{fig8} contains those
predicted by the NSF formulation. \ Comparing the curves in these two figures,
one observes that the mass density, temperature, and fluid velocity profiles
are much closer together in my theory than for the NSF formulation. \ This
difference is perhaps most obvious in the leading edge of the shock wave (the
$x^{\prime\prime}<0$ side) where my theory predicts these three profiles to be
very near one another, whereas NSF predicts a pronounced displacement between
the three with temperature leading, velocity behind it, and mass density in
the back. \ I am unfamiliar with any direct physical arguments or experimental
evidence that would explain why these profiles should be displaced in such a
manner. \ Therefore, it would be very interesting to obtain measurements
revealing the structure and position of these quantities. \ Experiments, such
as Alsmeyer's, which have been conducted to probe the internal structure of
shock waves in a gas, measure only the mass density profile.\footnote{When
comparing normalized mass density profiles from the $\overline{M}\left(
D,\eta\right)  $ and NSF formulations and Alsmeyer's curves fitted to
experimental values, one sees that both of the continuum formulations
underestimate the thickness of the shock wave. \ However, in view of the
concerns raised at the beginning of this section, it is unreasonable to expect
quantitative accuracy from a near-equilibrium continuum theory for this
problem. \ The development of non-linear constitutive equations may aid in
constructing a more realistic continuum model.}%

\section{Future Work}

To advance the study of single-component fluids with the $\overline{M}\left(
D,\eta\right)  $-formulation, there are two major tasks that lie ahead: (1) to
further investigate non-linear phenomena and (2) to reconcile the
$\overline{M}\left(  D,\eta\right)  $-description with macroscopic behavior
observed outside of the hydrodynamic regime, i.e. in the transition and
rarefied gas regimes. \ First, however, there remain a few important details
to address regarding the linearized equations in the hydrodynamic regime. \ I
treat these topics, mentioned below, in future papers.

\begin{itemize}
\item I recast the linearized $\overline{M}$-formulation (which includes both
the linearized $\overline{M}\left(  D,\eta\right)  $\ and NSF descriptions) in
the following way. \ Using Legendre transformations, I reexpress the equations
of motion in terms of variables, $\overline{m}$, $P$, $\underline{p}$,
$\sigma$.\footnote{Doing so, leads to convenience when using the $\overline
{M}\left(  D,\eta\right)  $-formulation since it gives rise to uncoupled
phenomena.} Afterwards, I normalize these variables in a way that facilitates
the study of hydrodynamical fluctuations, as treated in \cite{stochIII}.

\item For the recast linearized $\overline{M}$-formulation, I compute Green's
functions on one and three-dimensional infinite spatial domains. \ These are
then used to give a complete mathematical description of (1) the equilibration
of initial disturbances in an infinite fluid and (2) sound waves emanating
from a vibrating source into an infinite fluid.

\item The method of images is used to produce Green's functions for the
linearized $\overline{M}$-formulation on a one-dimensional semi-infinite
spatial domain with boundary conditions that prescribe%
\begin{equation}
\nabla P\cdot\underline{n}\text{, }\nabla\sigma\cdot\underline{n}\text{, and
}\underline{p}\cdot\underline{n}\label{cf1}%
\end{equation}
where $\underline{n}$\ represents the outward unit normal to the
boundary.\footnote{These types of conditions arise naturally in the
$\overline{M}\left(  D,\eta\right)  $-formulation when one wishes to specify
the normal mass and/or energy flux through a boundary.} \ These Green's
functions are employed to gain a complete mathematical description of
one-dimensional disturbances in a fluid interacting with an impermeable,
infinite impedance wall.

\item Green's functions are found for the linearized $\overline{M}%
$-formulation on a one-dimensional finite spatial domain with boundary
conditions of type (\ref{cf1}) in order to study sound resonators of infinite impedance.

\item The variables in the one-dimensional $\overline{M}\left(  D,\eta\right)
$-formulation are further recast in order to isolate the right-going and
left-going parts of the solution. \ Green's functions are found for this
recast system on infinite, semi-infinite, and finite spatial domains to use in
the study of fluid disturbances interacting with non-infinite impedance walls.

\item A detailed examination of stationary-Gaussian-Markov processes is
carried out in order to give a complete stochastical description of linearized
hydrodynamics with the $\overline{M}$-formulation. \ I divide this treatment
into three parts: I. general theory, II. Brownian oscillators\footnote{This is
an interlude that illustrates a simple concept analogous to the $\overline
{M}\left(  D,\eta\right)  $-formulation of hydrodynamics.} and, III.
hydrodynamical fluctuations.
\end{itemize}

Of course, there are eventually more complicated systems to be considered,
such as solids, multicomponent media, and electrically charged media. \ I wish
to treat all of these topics analogously to the way I have treated
single-component fluids. \ This will inevitably lead to descriptions that are
different from the ones currently used. \ As with the $\overline{M}\left(
D,\eta\right)  $-formulation for single-component fluids, I will ensure there
is full agreement with the standard models when physical, but there may also
arise disagreement in predictions--some testable by experiment and others
leading to new interpretations of phenomena.

\begin{center}
\textsc{Acknowledgement}
\end{center}

Well beyond the typical debt of gratitude a daughter owes her father, do I owe
mine, Marvin Morris. \ He has done literature searches, procured references,
and created the figures and graphs appearing in this paper, but most
importantly he has been my adviser in helping me map out a path through these studies.

\appendix

\section{\label{apptens}Tensors}

In this paper, I use the following identity:%

\begin{equation}
\nabla\cdot\left[  \left(  \nabla\underline{w}\right)  ^{sy,dev}\right]
=\frac{1}{2}\nabla^{2}\underline{w}+\frac{1}{6}\nabla\left(  \nabla
\cdot\underline{w}\right)  ,\label{to31r}%
\end{equation}
and cylindrical coordinate formulas:%
\begin{align}
\left(  \nabla\alpha\right)  _{r}  & =\frac{\partial\alpha}{\partial
r}\label{t1}\\
\left(  \nabla\alpha\right)  _{\theta}  & =\frac{1}{r}\frac{\partial\alpha
}{\partial\theta}\label{t2}\\
\left(  \nabla\alpha\right)  _{z}  & =\frac{\partial\alpha}{\partial
z}\label{t3}%
\end{align}%
\begin{equation}
\nabla^{2}\alpha=\frac{1}{r}\frac{\partial}{\partial r}\left(  r\frac
{\partial\alpha}{\partial r}\right)  +\frac{1}{r^{2}}\frac{\partial^{2}\alpha
}{\partial\theta^{2}}+\frac{\partial^{2}\alpha}{\partial z^{2}}\label{t4}%
\end{equation}%
\begin{equation}
\nabla\cdot\underline{w}=\frac{1}{r}\frac{\partial}{\partial r}\left(
rw_{r}\right)  +\frac{1}{r}\frac{\partial w_{\theta}}{\partial\theta}%
+\frac{\partial w_{z}}{\partial z}\label{t5}%
\end{equation}%
\begin{align}
\left(  \nabla^{2}\underline{w}\right)  _{r}  & =\frac{\partial}{\partial
r}\left[  \frac{1}{r}\frac{\partial}{\partial r}\left(  rw_{r}\right)
\right]  +\frac{1}{r^{2}}\frac{\partial^{2}w_{r}}{\partial\theta^{2}}%
+\frac{\partial^{2}w_{r}}{\partial z^{2}}-\frac{2}{r^{2}}\frac{\partial
w_{\theta}}{\partial\theta}\label{t6}\\
\left(  \nabla^{2}\underline{w}\right)  _{\theta}  & =\frac{\partial}{\partial
r}\left[  \frac{1}{r}\frac{\partial}{\partial r}\left(  rw_{\theta}\right)
\right]  +\frac{1}{r^{2}}\frac{\partial^{2}w_{\theta}}{\partial\theta^{2}%
}+\frac{\partial^{2}w_{\theta}}{\partial z^{2}}+\frac{2}{r^{2}}\frac{\partial
w_{r}}{\partial\theta}\label{t7}\\
\left(  \nabla^{2}\underline{w}\right)  _{z}  & =\frac{1}{r}\frac{\partial
}{\partial r}\left(  r\frac{\partial w_{z}}{\partial r}\right)  +\frac
{1}{r^{2}}\frac{\partial^{2}w_{z}}{\partial\theta^{2}}+\frac{\partial^{2}%
w_{z}}{\partial z^{2}}.\label{t8}%
\end{align}

\section{\label{appetr}Equilibrium Thermodynamic Relationships}

In addition to the thermodynamic parameters mentioned in \S 2 and
\textsc{appendix C}\ of \textsc{part I}, let us introduce the specific entropy
$\sigma$, isobaric specific heat per mass $c_{P}$, isobaric to isochoric
specific heat ratio $\gamma=c_{P}/c_{V}$, adiabatic sound speed $c$, the
particle number density,%
\begin{equation}
n_{p}=\frac{N_{A}}{A}m,\label{en1}%
\end{equation}
where $N_{A}$\ and $A$ respectively denote Avogadro's number and the atomic
weight, and the particle number $N_{p}=n_{p}V$. \ All of these quantities are
defined in Callen \cite{callen}.

For thermodynamically stable classical systems, in addition to (I.156) and
(I.157), the following inequalities hold:%
\begin{equation}
\sigma,c_{P},c>0,\label{e.4}%
\end{equation}
and%
\begin{equation}
\gamma\geq1.\label{e.5}%
\end{equation}

Note the following general equilibrium thermodynamic relationships:%
\begin{equation}
h_{M}=\frac{u+P}{m}\label{ee10}%
\end{equation}%
\begin{equation}
\kappa_{T}=\frac{\gamma}{mc^{2}}\label{e11}%
\end{equation}%
\begin{equation}
\frac{T\alpha_{P}}{m\kappa_{T}c_{V}}=\frac{\left(  \gamma-1\right)  }%
{\alpha_{P}}\label{e12}%
\end{equation}%
\begin{equation}
\gamma-1=\frac{T\alpha_{P}^{2}c^{2}}{c_{P}}\label{e12.1}%
\end{equation}%
\begin{align}
du  & =mc_{V}dT+\left(  h_{M}-\frac{T\alpha_{P}}{m\kappa_{T}}\right)
dm\label{e.6}\\
dP  & =\frac{\alpha_{P}}{\kappa_{T}}dT+\frac{1}{m\kappa_{T}}dm\label{e.7}%
\end{align}

\begin{align}
du  & =m\left(  c_{P}-h_{M}\alpha_{P}\right)  dT+\left(  \frac{h_{M}\gamma
}{c^{2}}-T\alpha_{P}\right)  dP\label{e4}\\
dm  & =-m\alpha_{P}dT+\frac{\gamma}{c^{2}}dP\label{e5}%
\end{align}
and%
\begin{equation}
d\sigma=\frac{c_{P}}{T}dT-\frac{\alpha_{P}}{m}dP\label{e5.5}%
\end{equation}
where equations (\ref{e.6})-(\ref{e5.5}) may be derived by using Legendre
transformations, as detailed in Callen \cite[\S 5.3]{callen}.

The equilibrium thermodynamic fluctuation formulas below may be found with the
methods in Callen \cite[\textsc{Ch. }15]{callen}. \ The following are second
moments that apply to a thermodynamic system of fixed volume $V$ in contact
with a heat/particle reservoir:%
\begin{align}
\left\langle \delta\!m^{2}\right\rangle _{V}  & =\frac{k_{B}m^{2}T\kappa_{T}%
}{V}\label{e40}\\
\left\langle \delta\!u^{2}\right\rangle _{V}  & =\frac{k_{B}mT}{V}\left[
Tc_{V}+m\kappa_{T}\left(  h_{M}-\frac{T\alpha_{P}}{m\kappa_{T}}\right)
^{2}\right] \label{e41}\\
\left\langle \delta\!m\delta\!u\right\rangle _{V}  & =\frac{k_{B}m^{2}%
T\kappa_{T}}{V}\left(  h_{M}-\frac{T\alpha_{P}}{m\kappa_{T}}\right)
\label{e42}\\
\left\langle \delta\!N_{p}^{2}\right\rangle _{V}  & =k_{B}n_{p}^{2}T\kappa
_{T}V\label{e42.1}%
\end{align}
where $k_{B}$\ denotes the Boltzmann constant.

Note the following classical ideal gas formulas:%
\begin{equation}
P=\frac{RTm}{A}\label{e25}%
\end{equation}%
\begin{equation}
\alpha_{P}=\frac{1}{T}\label{e26}%
\end{equation}%
\begin{equation}
\kappa_{T}=\frac{A}{RTm}\label{e27}%
\end{equation}%
\begin{equation}
c_{P}=\frac{R\gamma}{A\left(  \gamma-1\right)  }\label{e27.5}%
\end{equation}%
\begin{equation}
c=\sqrt{\frac{\gamma RT}{A}}\label{dc5}%
\end{equation}%
\begin{equation}
d\sigma=\frac{R\gamma}{A\left(  \gamma-1\right)  T}dT-\frac{R}{AP}dP\text{
(when }c_{V}\text{\ is constant)}\label{e28}%
\end{equation}
and classical monatomic ideal gas formulas:%
\begin{equation}
\gamma=\frac{5}{3}\label{e29}%
\end{equation}%
\begin{equation}
c=\sqrt{\frac{5RT}{3A}}\label{e30}%
\end{equation}%
\begin{equation}
u=\frac{3RTm}{2A}\label{e32}%
\end{equation}%
\begin{equation}
c_{V}=\frac{3R}{2A}\label{e33}%
\end{equation}
where $R$\ is the universal gas constant and $A$\ represents the atomic weight
of the gas.

\section{\label{difp}Values for $D$}

Here, measured sound propagation data is used together with formulas from
\S \ref{sound} to compute the longitudinal diffusion coefficient, $D$, of the
$\overline{M}\left(  D,\eta\right)  $-formulation for various types of fluids.
\ The temperature and pressure dependence of this parameter is explored where
data is available. \ In the future, I hope to obtain more (and more reliable)
data with which to refine these estimates and to include values for additional fluids.

\subsubsection*{Values at Normal Temperature and Pressure}

\textsc{Table} \ref{tab1} provides values of the $\overline{M}\left(
D,\eta\right)  $-formulation diffusion coefficient, $D_{\text{eq}}$,\ the
self-diffusion coefficient, $\left(  D_{\text{self}}\right)  _{\text{eq}}%
$,\ (where available), and the dimensionless parameter $C_{\text{eq}}$ defined
via equation (\ref{c35.5}), for several gases and liquids at $T_{\text{eq}%
}=300$ K (unless indicated otherwise) and $P_{\text{eq}}=1.013\times10^{5}$
Pa.%
\begin{table}[tbp] \centering
\caption{diffusion parameters for various fluids at normal temperature and
pressure}%
\begin{tabular}
[c]{|l|l|l|l|l|}\hline
\textbf{fluid} & $%
\begin{array}
[c]{c}%
D_{\text{eq}}\\
(\text{m}^{2}/\text{s})
\end{array}
$ & $%
\begin{array}
[c]{c}%
\left(  D_{\text{self}}\right)  _{\text{eq}}\\
(\text{m}^{2}/\text{s})
\end{array}
$ & $C_{\text{eq}}$ & $%
\begin{array}
[c]{c}%
\left(  \overline{m}D\right)  _{\text{eq}}\\
(\text{kg}/(\text{m}\cdot\text{s}))
\end{array}
$\\\hline
\textbf{gases} &  &  &  & \\\hline
helium & $1.44\times10^{-4}$ & $1.82\times10^{-4}$ & $1.17$ & $2.34\times
10^{-5}$\\\hline
neon & $4.57\times10^{-5}$ & $5.25\times10^{-5}$ & $1.17$ & $3.75\times
10^{-5}$\\\hline
argon & $1.64\times10^{-5}$ & $1.84\times10^{-5}$ & $1.17$ & $2.66\times
10^{-5}$\\\hline
krypton & $8.77\times10^{-6}$ & $9.84\times10^{-6}$ & $1.17$ & $2.99\times
10^{-5}$\\\hline
xenon & $5.07\times10^{-6}$ & $5.76\times10^{-6}$ & $1.17$ & $2.70\times
10^{-5}$\\\hline
nitrogen & $2.01\times10^{-5}$ & $2.12\times10^{-5}$ $^{\dag}$ & $1.28$ &
$2.29\times10^{-5}$\\\hline
oxygen & $1.79\times10^{-5}$ & $2.32\times10^{-5}$ $^{\dag}$ & $1.12$ &
$2.33\times10^{-5}$\\\hline
carbon dioxide & $1.08\times10^{-5}$ & $1.13\times10^{-5}$ $^{\dag}$ & $1.29$
& $1.93\times10^{-5}$\\\hline
methane & $2.59\times10^{-5}$ & $2.40\times10^{-5}$ $^{\dag}$ & $1.51$ &
$1.69\times10^{-5}$\\\hline
air & $1.96\times10^{-5}$ &  & $1.26$ & $2.17\times10^{-5}$\\\hline
\textbf{liquids} &  &  &  & \\\hline
water & $1.6\times10^{-6}$ $^{\ddag}$ & $2.60\times10^{-9}$ $^{\ddag}$ & $2.0$
$^{\ddag}$ & $1.6\times10^{-3}$ $^{\ddag}$\\\hline
mercury & $4.5\times10^{-7}$ $^{\ddag}$ & $5.7\times10^{-9}$ $^{\ddag}$ &
$4.1$ $^{\ddag}$ & $6.1\times10^{-3}$ $^{\ddag}$\\\hline
glycerol & $5.4\times10^{-4}$ $^{\dag}$ & $1.9\times10^{-12}$ $^{\dag}$ &
$0.78$ $^{\dag}$ & $6.8\times10^{-1}$ $^{\dag}$\\\hline
benzene & $4.8\times10^{-5}$ $^{\dag}$ & $2.2\times10^{-9}$ $^{\dag}$ & $70$
$^{\dag}$ & $4.22\times10^{-2}$ $^{\dag}$\\\hline
ethyl alcohol & $1.9\times10^{-6}$ $^{\dag}$ & $1.4\times10^{-9}$ $^{\dag}$ &
$1.4$ $^{\dag}$ & $1.5\times10^{-3}$ $^{\dag}$\\\hline
castor oil & $4.4\times10^{-4}$ $^{\dag}$ &  & $0.43$ $^{\dag}$ &
$4.2\times10^{-1}$ $^{\dag}$\\\hline
$%
\begin{array}
[c]{c}%
\dag\text{ at }T_{\text{eq}}=298\text{ K}\\
\ddag\text{ at }T_{\text{eq}}=303\text{ K}%
\end{array}
$ &  &  &  & \\\hline
\end{tabular}
\label{tab1}%
\end{table}
\ For the noble gases, helium, neon, argon, krypton, and xenon, $C_{\text{eq}%
}$\ is given by (\ref{c36}) and the diffusion coefficient is computed via
(\ref{c35}) with shear viscosities from Kestin et al.\ \cite{kestin}\ and mass
densities determined by the classical ideal gas relationship (\ref{e25}),
which is appropriate for each of the \textsc{table} \ref{tab1} gases in the
normal temperature and pressure regime. \ For the diatomic gases, nitrogen and
oxygen, and the polyatomic gases, carbon dioxide and methane, $D_{\text{eq}}$
and $C_{\text{eq}}$\ are computed by equations (\ref{c33}) and (\ref{c35.5}),
respectively, using data presented in Marques \cite{marques}, viscosity data
from Cole and Wakeham \cite{cole}\ and Trengrove and Wakeham \cite{tren}, and
mass densities computed via relation (\ref{e25}). \ For air\footnote{Note that
even though air is a mixture of gases, mainly nitrogen and oxygen, we can
treat it as though it were a single-component fluid as long as it remains a
homogeneous mixture--see \textsc{appendix D}\ of \textsc{part I}.} and the
liquids, $D_{\text{eq}}$ is computed via equation (\ref{c28.6}) and
$C_{\text{eq}}$ by relation (\ref{c35.5}). \ The sound propagation quantities,
$\alpha/f^{2}$ and $c_{\text{eq}}$, are calculated for air using
\cite{npl}\ with $f=11$ MHz\ and $0\%$\ humidity, given for mercury in Hunter
et al. \cite{hunter}, and tabulated for the rest of the liquids\ in
\cite{sounddata}. \ The viscosities and mass densities are taken from
\cite{airprop} for air, \cite{H2Odata} for water, \cite{hunter} for mercury,
\cite{glycdata}\ for glycerol, \cite{viscdata}\ and \cite{benzdata}\ for
benzene, \cite{ethdata} for ethyl alcohol, and \cite{viscdata}\ and
\cite{oildata}\ for castor oil. \ The self-diffusion coefficients are taken
from Kestin et al. \cite{kestin} for the noble gases, from Winn \cite{winn}
for the diatomic and polyatomic gases, from Holz et al. \cite{holz} for water,
from Nachtrieb and Petit \cite{nach}\ for mercury, from Tomlinson
\cite{tom}\ for glycerol, from Kim and Lee \cite{kim}\ for benzene, and from
Meckl and Zeidler \cite{meckl} for ethyl alcohol.

From the values presented in \textsc{table} \ref{tab1}, one makes the
following observations.

\begin{itemize}
\item In each of the gases, the diffusion and self-diffusion coefficients are
observed to be the same order of magnitude. \ However in liquids, the
self-diffusion coefficients are several orders of magnitude smaller than the
$D_{\text{eq}}$\ values. \ This indicates that although in gases the diffusion
coefficient\ may be roughly approximated by self-diffusion, the same may not
be said of liquids.

\item The diffusion parameters of the noble gases are seen to decrease with
increasing mass density.

\item The value for $D_{\text{eq}}$\ in air is approximately a weighted
average of the $D_{\text{eq}}$\ values for nitrogen and oxygen based on their
fractional compositions in air ($\approx0.8$ nitrogen and $\approx0.2 $ oxygen).

\item With the exception of very light gases like helium and very heavy gases
like krypton and xenon, diffusion coefficients for gases\ at normal
temperature and pressure are typically on the order of $10^{-5}$ m$^{2}/$s.

\item The $D_{\text{eq}}$\ values of water and ethyl alcohol--which fall into
the category of medium viscosity, medium density liquids--are similar and on
the order of $10^{-6}$ m$^{2}/$s, an order of magnitude smaller than that of a
typical gas. \ The high viscosity, medium density liquids, glycerol and castor
oil, possess similar $D_{\text{eq}}$\ values on the higher order of $10^{-4}$
m$^{2}/$s. \ Mercury's viscosity is medium range like water and ethyl alcohol,
but its much higher mass density results in a smaller $D_{\text{eq}}$\ on the
order of $10^{-7}$ m$^{2}/$s.

\item At normal temperature and pressure, all of the gases have $\left(
\overline{m}D\right)  _{\text{eq}}$ on the order of $10^{-5}$ kg/(m$\cdot$s)
with this quantity not varying much from gas to gas; and all of the liquids
have much higher values for $\left(  \overline{m}D\right)  _{\text{eq}}$ with
orders varying from $10^{-3}$ to $10^{-1}$\ kg/(m$\cdot$s) and the highest
values belonging to the most viscous fluids, glycerol and castor oil.
\end{itemize}

\subsubsection*{Dependence on State Parameters}

\paragraph*{\textbf{Gases}}

In general, it is observed from sound attenuation data for gases in the
hydrodynamic regime that the product, $\left(  \overline{m}D\right)
_{\text{eq}}=m_{\text{eq}}D_{\text{eq}}$, may vary with temperature but has
negligible pressure dependence at a fixed temperature. \ Using this
observation, together with relationship (I.72)\ and the fact that the shear
viscosity $\eta$\ also has negligible pressure dependence, leads us to
conclude that, although $C$ is possibly a function of $T$, it does not tend to
vary much with $P$\ for gases. \ Typically, $\eta$\ may be expressed as a
temperature power law of the form,%
\begin{equation}
\eta=\eta_{r}\left(  \frac{T}{T_{r}}\right)  ^{\beta},\label{dc1}%
\end{equation}
where $T_{r}$\ is some reference temperature and $\eta_{r}$\ is the viscosity
measured at $T_{r}$. \ Using this in equation (I.7.7), one finds%
\begin{equation}
\left(  \overline{m}D\right)  \left(  T\right)  =\eta_{r}C\left(  T\right)
\left(  \frac{T}{T_{r}}\right)  ^{\beta}.\label{dc2}%
\end{equation}

Equations (\ref{c33}) and (\ref{c35.5}) together imply%
\begin{equation}
C_{\text{eq}}=\frac{1}{2}\left\{  \frac{\zeta_{NS}}{\eta}+\left[  \frac{4}%
{3}+\left(  1-\frac{1}{\gamma}\right)  E\!u\right]  \right\}  _{\text{eq}%
}.\label{dc3}%
\end{equation}
For noble gases, such as argon, one expects the values for $\left(  \zeta
_{NS}\right)  _{\text{eq}}$, $\gamma_{\text{eq}}$, and $E\!u_{\text{eq}}%
$\ given in (\ref{c34}) to hold over a broad temperature range. \ Therefore,
in the noble gases, we take $C=7/6$\ to be constant and, substituting this
into (\ref{dc2}),%
\begin{equation}
\left(  \overline{m}D\right)  \left(  T\right)  =\frac{7}{6}\eta_{r}\left(
\frac{T}{T_{r}}\right)  ^{\beta}.\label{dc4}%
\end{equation}
On the other hand, as observed in \textsc{tables} \ref{tab2}-\ref{tab3.5}, for
non-monatomic gases such as air, nitrogen, and methane, $C$\ displays a
tendency to increase with temperature. \ In \textsc{tables} \ref{tab2}%
-\ref{tab3.5}, the $D_{\text{eq}}$\ values are computed by equation
(\ref{c28.6}) with $\alpha/f^{2}$ calculated using \cite{npl}\ for air with
$f=11$ MHz\ and $0\%$\ humidity and taken from \cite{prang}\ for nitrogen and
methane and with $c_{\text{eq}}$\ computed from the classical ideal gas
formula (\ref{dc5}), where the $\gamma_{\text{eq}}$\ values are taken from
\cite{airprop}\ for air and approximated to be constant over the studied
temperature range for nitrogen and methane with respective values $1.4$\ and
$1.3$. \ The $C_{\text{eq}}$\ values appearing in \textsc{tables}
\ref{tab2}-\ref{tab3.5} are computed via equation (\ref{c35.5}) with shear
viscosities taken from \cite{airprop}\ for air and from \cite{prang}\ for
nitrogen and methane and mass densities taken from \cite{airprop}\ for air and
computed by ideal gas formula (\ref{e25}) for nitrogen and methane.%

\begin{table}[tbp] \centering
\caption{diffusion coefficient for air at atmospheric pressure and various
temperatures}%
\begin{tabular}
[c]{|l|l|l|l|}\hline
$T_{\text{eq}}$ (K) & $D_{\text{eq}}$ (m$^{2}/$s) & $C_{\text{eq}}$ & $\left(
\overline{m}D\right)  _{\text{eq}}$ (kg/(m$\cdot$s))\\\hline
$253$ & $1.40\times10^{-5}$ & $1.22$ & $1.96\times10^{-5}$\\\hline
$263$ & $1.52\times10^{-5}$ & $1.23$ & $2.04\times10^{-5}$\\\hline
$273$ & $1.64\times10^{-5}$ & $1.24$ & $2.11\times10^{-5}$\\\hline
$283$ & $1.76\times10^{-5}$ & $1.25$ & $2.19\times10^{-5}$\\\hline
$288.15$ & $1.82\times10^{-5}$ & $1.25$ & $2.23\times10^{-5}$\\\hline
$293$ & $1.88\times10^{-5}$ & $1.26$ & $2.27\times10^{-5}$\\\hline
$303$ & $2.01\times10^{-5}$ & $1.27$ & $2.35\times10^{-5}$\\\hline
$313$ & $2.15\times10^{-5}$ & $1.27$ & $2.42\times10^{-5}$\\\hline
$323$ & $2.29\times10^{-5}$ & $1.28$ & $2.50\times10^{-5}$\\\hline
\end{tabular}
\label{tab2}%
\end{table}%
%

\begin{table}[tbp] \centering
\caption{diffusion coefficient for nitrogen gas at 0.01 atm and various
temperatures}%
\begin{tabular}
[c]{|l|l|l|l|}\hline
$T_{\text{eq}}$ (K) & $D_{\text{eq}}$ (m$^{2}/$s) & $C_{\text{eq}}$ & $\left(
\overline{m}D\right)  _{\text{eq}}$ (kg/(m$\cdot$s))\\\hline
$77.1$ & $1.37\times10^{-4}$ & $1.11$ & $6.09\times10^{-6}$\\\hline
$180$ & $7.63\times10^{-4}$ & $1.22$ & $1.42\times10^{-5}$\\\hline
$260$ & $1.57\times10^{-3}$ & $1.29$ & $2.06\times10^{-5}$\\\hline
$293$ & $1.97\times10^{-3}$ & $1.31$ & $2.32\times10^{-5}$\\\hline
\end{tabular}
\label{tab3}%
\end{table}%
%

\begin{table}[tbp] \centering
\caption{diffusion coefficient for methane gas at 0.01 atm and various
temperatures}%
\begin{tabular}
[c]{|l|l|l|l|}\hline
$T_{\text{eq}}$ (K) & $D_{\text{eq}}$ (m$^{2}/$s) & $C_{\text{eq}}$ & $\left(
\overline{m}D\right)  _{\text{eq}}$ (kg/(m$\cdot$s))\\\hline
$77.1$ & $1.58\times10^{-4}$ & $1.26$ & $4.00\times10^{-6}$\\\hline
$180$ & $8.84\times10^{-4}$ & $1.36$ & $9.79\times10^{-6}$\\\hline
$260$ & $1.92\times10^{-3}$ & $1.47$ & $1.44\times10^{-5}$\\\hline
$293$ & $2.49\times10^{-3}$ & $1.52$ & $1.64\times10^{-5}$\\\hline
\end{tabular}
\label{tab3.5}%
\end{table}
\ The following are least squares fits to the data appearing in
\textsc{tables} \ref{tab2}-\ref{tab3.5}:%
\begin{align}
\left(  \overline{m}D\right)  _{\text{eq}}  & =2.32\times10^{-5}\left[
\frac{T_{\text{eq}}\text{ }\left(  \text{in K}\right)  }{300}\right]
^{1.00}\text{ kg/(m}\cdot\text{s) for air}\label{dc4.1}\\
\left(  \overline{m}D\right)  _{\text{eq}}  & =2.38\times10^{-5}\left[
\frac{T_{\text{eq}}\text{ }\left(  \text{in K}\right)  }{300}\right]
^{1.00}\text{ kg/(m}\cdot\text{s) for N}_{2}\text{(g)}\label{dc4.2}\\
\left(  \overline{m}D\right)  _{\text{eq}}  & =1.68\times10^{-5}\left[
\frac{T_{\text{eq}}\text{ }\left(  \text{in K}\right)  }{300}\right]
^{1.05}\text{ kg/(m}\cdot\text{s) for CH}_{4}\text{(g).}\label{dc4.3}%
\end{align}

\paragraph*{\textbf{Liquids}}

Using equations (\ref{c28.6}) and (\ref{c35.5}) with sound speed and
attenuation data from \cite{sounddata} and mass density and shear viscosity
data from \cite{H2Odata}, one obtains the estimates of $D_{\text{eq}}$\ and
$C_{\text{eq}}$ for water at atmospheric pressure ($P_{\text{eq}}%
=1.013\times10^{5}$ Pa) and various temperatures $T_{\text{eq}}$ between the
freezing and boiling point appearing in \textsc{table} \ref{tab4}.%
\begin{table}[tbp] \centering
\caption{diffusion parameters for water at atmospheric pressure and various
temperatures between the freezing and boiling point}%
\begin{tabular}
[c]{|c|c|c|c|c|}\hline
$T_{\text{eq}}$ (K) & $D_{\text{eq}}$ (m$^{2}$/s) & $\left(  D_{\text{self}%
}\right)  _{\text{eq}}$ (m$^{2}$/s) & $C_{\text{eq}}$ & $\left(  \overline
{m}D\right)  _{\text{eq}}$ (kg/(m$\cdot$s))\\\hline
$273$ & $4.0\times10^{-6}$ & $1.10\times10^{-9}$ & $2.2$ & $4.0\times10^{-3}%
$\\\hline
$283$ & $2.8\times10^{-6}$ & $1.53\times10^{-9}$ & $2.1$ & $2.8\times10^{-3}%
$\\\hline
$293$ & $2.1\times10^{-6}$ & $2.02\times10^{-9}$ & $2.1$ & $2.1\times10^{-3}%
$\\\hline
$303$ & $1.6\times10^{-6}$ & $2.59\times10^{-9}$ & $2.0$ & $1.6\times10^{-3}%
$\\\hline
$313$ & $1.3\times10^{-6}$ & $3.24\times10^{-9}$ & $2.0$ & $1.3\times10^{-3}%
$\\\hline
$323$ & $1.1\times10^{-6}$ & $3.96\times10^{-9}$ & $2.0$ & $1.1\times10^{-3}%
$\\\hline
$333$ & $9.5\times10^{-7}$ & $4.75\times10^{-9}$ & $2.0$ & $9.3\times10^{-4}%
$\\\hline
$343$ & $8.3\times10^{-7}$ & $5.62\times10^{-9}$ & $2.0$ & $8.1\times10^{-4}%
$\\\hline
$353$ & $7.5\times10^{-7}$ & $6.56\times10^{-9}$ & $2.1$ & $7.3\times10^{-4}%
$\\\hline
$363$ & $6.9\times10^{-7}$ & $7.57\times10^{-9}$ & $2.1$ & $6.7\times10^{-4}%
$\\\hline
$373$ & $6.5\times10^{-7}$ & $8.67\times10^{-9}$ & $2.2$ & $6.2\times10^{-4}%
$\\\hline
\end{tabular}
\label{tab4}%
\end{table}
\ As one can see, the parameter $C_{\text{eq}}$\ does not vary much over the
entire temperature range, but $D_{\text{eq}}$\ and $\left(  \overline
{m}D\right)  _{\text{eq}}$\ decrease with temperature. \ Excellent least
squares fits are made to the water data in \textsc{table} \ref{tab4} with%
\begin{equation}
D_{\text{eq}}=3.9853\times10^{-7}\exp\left(  \frac{3.5115\times10^{12}%
}{T_{\text{eq}}^{5}}\right)  \text{ m}^{2}\text{/s}\label{dc6}%
\end{equation}
and%
\begin{equation}
\left(  \overline{m}D\right)  _{\text{eq}}=3.8262\times10^{-4}\exp\left(
\frac{3.5899\times10^{12}}{T_{\text{eq}}^{5}}\right)  \text{ kg/(m}%
\cdot\text{s)}\label{dc6.1}%
\end{equation}
where $T_{\text{eq}}$ is in Kelvin. \ In the third column of \textsc{table}
\ref{tab4}, measured self-diffusion coefficients from Holz et al.
\cite{holz}\ are tabulated for water at atmospheric pressure. \ Unlike our
observation for gases, these self-diffusions are two to three orders of
magnitude less than the corresponding $D_{\text{eq}}$ values. \ Also, unlike
$D_{\text{eq}}$, the self-diffusion \textit{increases} with temperature.

In \textsc{table} \ref{tab5}, $D_{\text{eq}}$\ and $C_{\text{eq}}$\ are
computed for water at two fixed temperatures, $273$ K\ and $303$ K,\ and
various pressures roughly between $1$\ and $2000$\ atmospheres. The values are
obtained using equations (\ref{c28.6}) and (\ref{c35.5}) together with data
presented in Litovitz and Carnevale \cite{lito}. \
\begin{table}[tbp] \centering
\caption{diffusion coefficient for water at 273 K and 303 K and various
pressures}%
\begin{tabular}
[c]{|llll|}\hline
$P_{\text{eq}}$ (Pa) & \multicolumn{1}{|l}{$D_{\text{eq}}$ (m$^{2}$/s)} &
\multicolumn{1}{|l}{$C_{\text{eq}}$} & \multicolumn{1}{|l|}{$\left(
\overline{m}D\right)  _{\text{eq}}$ (kg/(m$\cdot$s))}\\\hline\hline
$T_{\text{eq}}=273$ K &  &  & \\\hline\hline
$9.81\times10^{4}$ & \multicolumn{1}{|l}{$4.03\times10^{-6}$} &
\multicolumn{1}{|l}{$2.25$} & \multicolumn{1}{|l|}{$4.03\times10^{-3}$%
}\\\hline
$4.90\times10^{7}$ & \multicolumn{1}{|l}{$3.96\times10^{-6}$} &
\multicolumn{1}{|l}{$2.43$} & \multicolumn{1}{|l|}{$4.08\times10^{-3}$%
}\\\hline
$9.81\times10^{7}$ & \multicolumn{1}{|l}{$3.85\times10^{-6}$} &
\multicolumn{1}{|l}{$2.42$} & \multicolumn{1}{|l|}{$3.93\times10^{-3}$%
}\\\hline
$1.47\times10^{8}$ & \multicolumn{1}{|l}{$3.59\times10^{-6}$} &
\multicolumn{1}{|l}{$2.26$} & \multicolumn{1}{|l|}{$3.77\times10^{-3}$%
}\\\hline
$1.96\times10^{8}$ & \multicolumn{1}{|l}{$3.39\times10^{-6}$} &
\multicolumn{1}{|l}{$2.13$} & \multicolumn{1}{|l|}{$3.73\times10^{-3}$%
}\\\hline\hline
$T_{\text{eq}}=303$ K &  &  & \\\hline\hline
$9.81\times10^{4}$ & \multicolumn{1}{|l}{$1.61\times10^{-6}$} &
\multicolumn{1}{|l}{$2.02$} & \multicolumn{1}{|l|}{$1.60\times10^{-3}$%
}\\\hline
$4.90\times10^{7}$ & \multicolumn{1}{|l}{$1.58\times10^{-6}$} &
\multicolumn{1}{|l}{$1.93$} & \multicolumn{1}{|l|}{$1.56\times10^{-3}$%
}\\\hline
$9.81\times10^{7}$ & \multicolumn{1}{|l}{$1.52\times10^{-6}$} &
\multicolumn{1}{|l}{$1.84$} & \multicolumn{1}{|l|}{$1.60\times10^{-3}$%
}\\\hline
$1.47\times10^{8}$ & \multicolumn{1}{|l}{$1.52\times10^{-6}$} &
\multicolumn{1}{|l}{$1.76$} & \multicolumn{1}{|l|}{$1.58\times10^{-3}$%
}\\\hline
$1.96\times10^{8}$ & \multicolumn{1}{|l}{$1.54\times10^{-6}$} &
\multicolumn{1}{|l}{$1.82$} & \multicolumn{1}{|l|}{$1.60\times10^{-3}$%
}\\\hline
\end{tabular}
\label{tab5}%
\end{table}%
As one can see in the fourth column of \textsc{table} \ref{tab5}, $\left(
\overline{m}D\right)  _{\text{eq}}$ does not vary much over the studied
pressure range ($8.96\%$ for $273$ K and $2.53\%$ for $303$ K). \ Therefore,
as in gases, it is reasonable to assume that the quantity $\overline{m}D$ is
essentially pressure-independent for water and perhaps for other liquids, as
well, although this remains to be experimentally verified.

Using equations (\ref{c28.6}) and (\ref{c35.5}) with data from Hunter et al.
\cite{hunter} yields the estimates of $D_{\text{eq}}$\ and $C_{\text{eq}}$ for
liquid mercury at atmospheric pressure ($P_{\text{eq}}=1.013\times10^{5} $ Pa)
and various temperatures $T_{\text{eq}}$ between $298$ K and $403$ K appearing
in \textsc{table} \ref{tab6}. \
\begin{table}[tbp] \centering
\caption{diffusion coefficient for liquid mercury at atmospheric pressure
and various temperatures}%
\begin{tabular}
[c]{|l|l|l|l|}\hline
$T_{\text{eq}}$ (K) & $D_{\text{eq}}$ (m$^{2}/$s) & $C_{\text{eq}}$ & $\left(
\overline{m}D\right)  _{\text{eq}}$ (kg/(m$\cdot$s))\\\hline
$298$ & $4.41\times10^{-7}$ & $3.90$ & $5.97\times10^{-3}$\\\hline
$303$ & $4.46\times10^{-7}$ & $4.05$ & $6.03\times10^{-3}$\\\hline
$313$ & $4.63\times10^{-7}$ & $4.31$ & $6.25\times10^{-3}$\\\hline
$318$ & $4.76\times10^{-7}$ & $4.51$ & $6.41\times10^{-3}$\\\hline
$333$ & $5.04\times10^{-7}$ & $5.01$ & $6.77\times10^{-3}$\\\hline
$338$ & $5.15\times10^{-7}$ & $5.20$ & $6.92\times10^{-3}$\\\hline
$343$ & $5.19\times10^{-7}$ & $5.32$ & $6.97\times10^{-3}$\\\hline
$353$ & $5.34\times10^{-7}$ & $5.64$ & $7.16\times10^{-3}$\\\hline
$378$ & $5.74\times10^{-7}$ & $6.27$ & $7.65\times10^{-3}$\\\hline
$403$ & $6.10\times10^{-7}$ & $6.92$ & $8.10\times10^{-3}$\\\hline
\end{tabular}
\label{tab6}%
\end{table}%
Unlike our observation for water in \textsc{table} \ref{tab4}, the
$D_{\text{eq}}$\ and $\left(  \overline{m}D\right)  _{\text{eq}}$\ values for
mercury are seen to increase with increasing temperature. \ The following are
least squares fits to the mercury data in \textsc{table} \ref{tab6}:%
\begin{equation}
D_{\text{eq}}=4.46\times10^{-7}\left[  \frac{T_{\text{eq}}\text{ }\left(
\text{in K}\right)  }{300}\right]  ^{1.10}\text{ m}^{2}\text{/s}\label{dc7}%
\end{equation}
and%
\begin{equation}
\left(  \overline{m}D\right)  _{\text{eq}}=6.03\times10^{-3}\left[
\frac{T_{\text{eq}}\text{ }\left(  \text{in K}\right)  }{300}\right]
^{1.04}\text{ kg/(m}\cdot\text{s).}\label{dc7.1}%
\end{equation}

\end{document}